\documentclass[manuscript]{aastex}

\slugcomment{Draft Copy \today}

\shorttitle{Methanol Masers}
\shortauthors{G\'omez et al.}

\begin{document}
\title{A Catalog of Methanol Masers in Massive Star-forming Regions.
\\III. The Molecular Outflow Sample}

\author{A. I. Gomez-Ruiz\altaffilmark{1,2,3}}
\affil{Instituto Nacional de Astrof\'isica, \'Optica y Electr\'onica, Luis E. Erro 1, Tonantzintla, Puebla, M\'exico, C.P. 72840}

\author{S. E. Kurtz}
\affil{Instituto de Radioastronom\'\i a y Astrof\'\i sica, Universidad Nacional Aut\'onoma de M\'exico, Apartado Postal 3-72, Morelia 58089, M\'exico}

\author{E. D. Araya}
\affil{Physics Department, Western Illinois University, 1 University Circle, Macomb, IL 61455, USA}

\author{P. Hofner\altaffilmark{4} }
\affil{New Mexico Institute of Mining and Technology, Socorro, NM 87801, USA}

\and

\author{L. Loinard\altaffilmark{5}}
\affil{Instituto de Radioastronom\'\i a y Astrof\'\i sica, Universidad Nacional Aut\'onoma de M\'exico, Apartado Postal 3-72, Morelia 58089, M\'exico}

\altaffiltext{1}{Consejo Nacional de Ciencia y Tecnolog\'ia, Av. Insurgentes Sur 1582, C.P. 03940, M\'exico}
\altaffiltext{2}{Instituto de Radioastronom\'\i a y Astrof\'\i sica, UNAM,Apartado Postal 3-72, Morelia 58089, M\'exico}
\altaffiltext{3}{C\'atedra CONACYT para J\'ovenes Investigadores}
\altaffiltext{4}{Adjunct Astronomer at the National Radio Astronomy Observatory}
\altaffiltext{5}{Max-Planck-Institut f\"ur Radioastronomie, Auf dem H\"ugel 69, 53121 Bonn, Germany}

\begin{abstract}

We present an interferometric survey of the 44 GHz class I methanol
maser transition toward a sample of 69 sources consisting of High Mass
Protostellar Object candidates and Ultracompact (UC) H{\,\small II}
regions. We found a 38\% detection rate (16 of 42) in the HMPO candidates and a
54\% detection rate (13 of 24) for the regions with ionized gas. This result
indicates that class I methanol maser emission is more common toward
more evolved young stellar objects of our sample. Comparing with similar interferometric data sets, our observations show narrower linewidths, likely due to our higher spatial resolution. Based on a comparison between
molecular outflow tracers and the maser positions, we find several
cases where the masers appear to be located at the outflow interface
with the surrounding core.   
Unlike previous surveys, we also find several cases where
the masers appear to be located close to the base of the molecular
outflow, although we can not discard projection effects. This and other surveys of class I methanol masers not only suggest that these masers may trace shocks at different stages, but may even trace shocks arising from a number of different phenomena occurring in star-forming regions: young/old outflows, cloud-cloud collisions, expanding H\,\small II regions, among others.

\end{abstract}


\section{Introduction}

Massive star formation begins with cold massive gas cores and ends
with the ionization of the circumstellar gas by the newly formed
massive stars \citep[e.g.,][]{Beuther07}. Hot molecular cores (HMCs),
molecular outflows and maser emission of different molecular species
are common phenomena during the early stages of high-mass star formation. Sources presenting these phenomena are generally referred to as High Mass Protostellar Objects (HMPOs) and typically correspond to objects in stages prior to the formation of ionized regions [Hypercompact (HC) or Ultracompact (UC) H{\,\small II regions].

Methanol (CH$_3$OH) is one of the molecular species that often shows
maser emission in high-mass star-forming regions.  Maser emission from
this molecule has been classified into two types, denoted as class I
and class II. The original classification scheme was based on the
association of maser emission with other star formation phenomena, and
with spectral characteristics such as the presence/absence of
other CH$_3$OH transitions and spectral complexity (Menten 1991). Class I methanol masers are usually observed offset 
 (between $\sim$ 0.1 to 1 pc) from UC HII
regions, OH and H$_2$O masers, and far
infrared (FIR) sources. These masers are often found in
association with high-velocity outflows and cloud-cloud collisions;
 collisional excitation has been proposed as their pumping
mechanism (Menten, 1996; Cragg et al. 1992). On the other
hand, class II methanol masers are usually found in close proximity to
young massive stellar objects (embedded sources), suggesting radiative excitation as the
pumping mechanism (Cragg et al. 2005, 1992). Class I masers have relatively
simple spectra with distinct velocity  components 
distributed over a narrower velocity range than H$_2$O masers, while Class II methanol
masers have a velocity range similar to that of OH masers and the
complexity of their spectra is similar to those of H$_2$O (Menten, 1991).

One of the most widespread and strongest class I methanol masers is
the 7$_0 \rightarrow$ 6$_1$ $A^+$ transition at 44 GHz
\cite{Bachiller90,Haschick90}. In recent years a number of
  interferometric observations, mainly obtained with the VLA and ATCA, have
  provided a better understanding of the spatial distribution of
  class I masers with respect to other star-formation
  tracers.  Kurtz et al. (2004) and G\'omez et al. (2010) (Papers I and
  II) observed a total 47 massive star-forming regions that were
  selected to include well-known regions spanning from early to evolved phases of
  evolution. Their surveys showed a large number of these masers in
  relatively close association ($\lesssim 0.3$ pc) with H\,\small II regions and water
  masers. Cyganowski et al. (2009), targeting 20 massive YSOs
  candidates with outflows identified from the excess in the 4.5
  micron IRAC/Spitzer band (Extended Green Objects, EGOs), found that
  most of the masers were located in close proximity to EGOs and
  therefore closely related with young and energetic outflows. More
  recently, Voronkov et al. (2014) presented 44 GHz and 36 GHz class I
  methanol maser observations of 71 southern sources with ATCA, noting
  that class I masers are typically related with outflows, expanding HII
  regions, dark clouds, shocks traced by the 4.5 micron excess, and 8
  micron filaments. 

This paper is the third in a series, presenting interferometric observations of the 44 GHz methanol maser in northern high-mass star-forming regions. The present paper covers a sub-sample of Young Stellar Objects (YSOs) reported by Molinari et al. (1996). In this paper we focus on the maser detection rates within this sub-sample, the line properties, a detailed comparison with the available cm-continuum and molecular outflows tracers, a color-color and far-IR properties analysis using the infrared data that originally defined this sample (i.e., IRAS), and finally provide a comparison with results from other authors with similar data sets (e.g., Voronkov et al. 2014). A forthcoming paper (Rodr\'\i guez-Garza et al.; in prep.) will present observations of a complementary sample of massive YSOs identified by Sridharan et al. (2002). A final paper in the series will present a complete analysis of the entire survey data, including an study of the infrared emission (in multiple bands) from the central objects using more recent infrared data (i.e., Spitzer, WISE, and Herschel).

\section{The sample}

Several large samples of young massive star formation sites have been
reported in the literature. Particularly notable is the sample of
Molinari et al. (1996; hereafter M96), 
since it has been followed-up in many different tracers, covering a wide 
range of scales and physical conditions}. The
M96 catalog consists of 163 sources selected from the water maser
survey of Palla et al. (1991). Sources in this catalog have IRAS
colors typical of compact molecular clouds and/or UC H\,\small II
regions. Sources with colors typical of UC H\,\small II
regions were termed {\it High} while the remaining sources were termed
{\it Low}. A subset of 69 sources, including both {\it High} (32) and
{\it Low} (37), was observed in CO by Zhang et al. (2001, 2005), who found outflow activity in 54\% of the sample. In this paper we present VLA
observations of the 44 GHz methanol transition toward this sub-sample
of 69 sources, thus complementing the survey of papers I and II by adding significantly more sources with outflows. We point out that there are a few sources in common among these samples. These common sources were re-observed in order to have a uniform data set within each sample.

Follow-up observations by Molinari et al. (1998; hereafter M98) showed that the {\it Low} group contains
objects in two different evolutionary stages: both with and without UC
HII regions.  This is problematic if we want to use {\it High} and
{\it Low} as an evolutionary stage indicator. To provide a more
definite age criterion, we use the presence or absence of centimeter
continuum emission as found in Very Large Array observations reported
by M98, Palau et al. (2010), and Kurtz, Churchwell,
\& Wood (1994).  M98 did not report observations for Mol 2
or Mol 15; for these two sources we use the data of Palau et al. (2010) and
Kurtz et al. (2004), respectively. In some cases the reported values are as low as 1
mJy. In what follows, we will use the phrase ``UC HII candidate region'' to
refer to those sources showing centimeter continuum emission at levels
$>$ 1 mJy.  We caution, however, that in some cases the continuum emission, apart from distance effects, may arise from some other process than photo-ionization of the medium surrounding an OB-type star. A number of different phenomena can give rise to centimeter continuum at this level; see, for example, Rodr\'\i guez et al. (2012).

Following M98, we consider that the IRAS and the centimeter continuum sources are associated if the peak of the continuum emission is within 40$''$ of the IRAS position. In Table 1 we list the sample of 69 sources, indicating whether they were considered {\it High} or {\it Low} by M96, and if they are associated with centimeter continuum emission and/or outflows. In the final column of Table 1, a $U$ indicates that  centimeter continuum emission has been reported within 40$''$ of the IRAS position; we consider these 25 sources to be our ``UC HII candidate sample''.  The 44 sources for which no $U$ is present form our ``HMPO candidate sample''.  The final column of Table 1 also shows the letter $O$ for those sources in which Zhang et al. (2005) reported a molecular outflow.

\section{Observations and data reduction}

Observations were made with the VLA of the NRAO\footnote{The National Radio Astronomy Observatory is a facility of the National Science Foundation operated under cooperative agreement by Associated Universities, Inc.} in the D configuration in five observing runs on 2007 March 22 and 30, and 2007 April 2, 20 and 30. Observations were centered on the CH$_3$OH (7$_0$-$6_1$) transition ($\nu_0 =$44.069430 GHz). A total of 127 channels were used in a 3.13 MHz bandwidth for the methanol line, giving a resolution of 0.17 km s$^{-1}$ and a velocity coverage of 21 km s$^{-1}$. The pointing center for each source was the position of the IRAS source and the band central velocity was taken from the ammonia observations of M96 (see Table 1). The observation mode was fast-switching, with a typical time on source of 8$-$10 min. Referenced pointing was performed approximately every hour or when changing to a new phase calibrator. No bandpass calibration was applied. 

The data were calibrated and imaged using the AIPS software package. Self-calibration was performed for sources with sufficient signal-to-noise ratio. For each source the strongest and most-isolated maser was selected as the model for the phase self-calibration. When appropriate, a second iteration in phase and amplitude was performed. The image cubes were made by setting CLEAN boxes for all maser components and CLEANing to a flux cutoff of four times the theoretical noise level. The average synthesized beam was 2.0$''$$\times$1.5$''$. Sources with low signal-to-noise (which were not self-calibrated) were imaged in the same way.

The observations were made during the Karl G. Jansky VLA upgrade, so on-line Doppler tracking was no longer available. We observed using absolute sky frequencies, and used the NRAO DOPSET program to calculate the corresponding LSR velocities, which were inserted into the image data headers during post-processing.

Table 1 lists the 69 sources observed. The table gives the Molinari and IRAS numbers, the J2000 coordinates of the IRAS source, the central velocity observed, synthesized beam size, line-free channel map rms noise and the detection result.  The latter value is the number of distinct maser components that we identify within the field. Finally, the ninth column indicates if an ionized gas region was detected in the observations of M98 or if a molecular outflow was reported by Zhang et al. (2005).  Mol 136, Mol 138 and Mol 139 were not calibrated or imaged because their phase calibrator was too weak. Mol 136 and Mol 138 were observed by Kurtz et al. (2004), who reported 44 GHz masers in both sources.  Nevertheless, we exclude all three sources from our statistical analysis.

\section{Results} \label{results}

We detected 44 GHz methanol maser emission in 29 of the 66 mapped
sources (44 \% detection rate). A total of 102 distinct maser
components were found in the survey. The parameters of these masers
are presented in Table 2. Column (1) gives the IRAS name and Molinari
number of the field, column (2) specifies the number of the maser
component assigned in order of increasing right ascension, columns (3)
and (4) are the positions of the component in $\alpha$(J2000) and
$\delta$(J2000), respectively, obtained by a two-dimensional (2D) spatial 
Gaussian fit to the peak channel map, column (5) is the V$_{LSR}$ of the peak
channel, column (6) gives the peak flux density from the 2D Gaussian fit of
the peak channel image, and column (7) is the full width at zero intensity
(FWZI) of the maser line defined by those channels above a
4$\sigma$ intensity level. When multiple velocity components are found
at the same position, we give the total velocity range. Column (8)
gives the velocity-integrated flux of each maser component. Eleven of
the 102 masers were found to spatially overlap a stronger maser
component and hence their positions are not well-determined. These 
overlapping components are indicated by footnotes in
column (9).

The strongest maser in our sample showed a flux density of
$580\,$Jy, and was detected toward IRAS$\,$18018$-$2426
(Mol$\,$37). This maser is among the strongest 44 GHz CH$_3$OH masers
ever detected (e.g., Kogan \& Slysh 1998; Jordan et al. 2008; Pratap
et al. 2008; Fontani et al. 2010).

\section{Discussion}

\subsection{Line properties}

\subsubsection{Linewidths}

\begin{figure}
\begin{center}
\includegraphics[angle=90,scale=0.65]{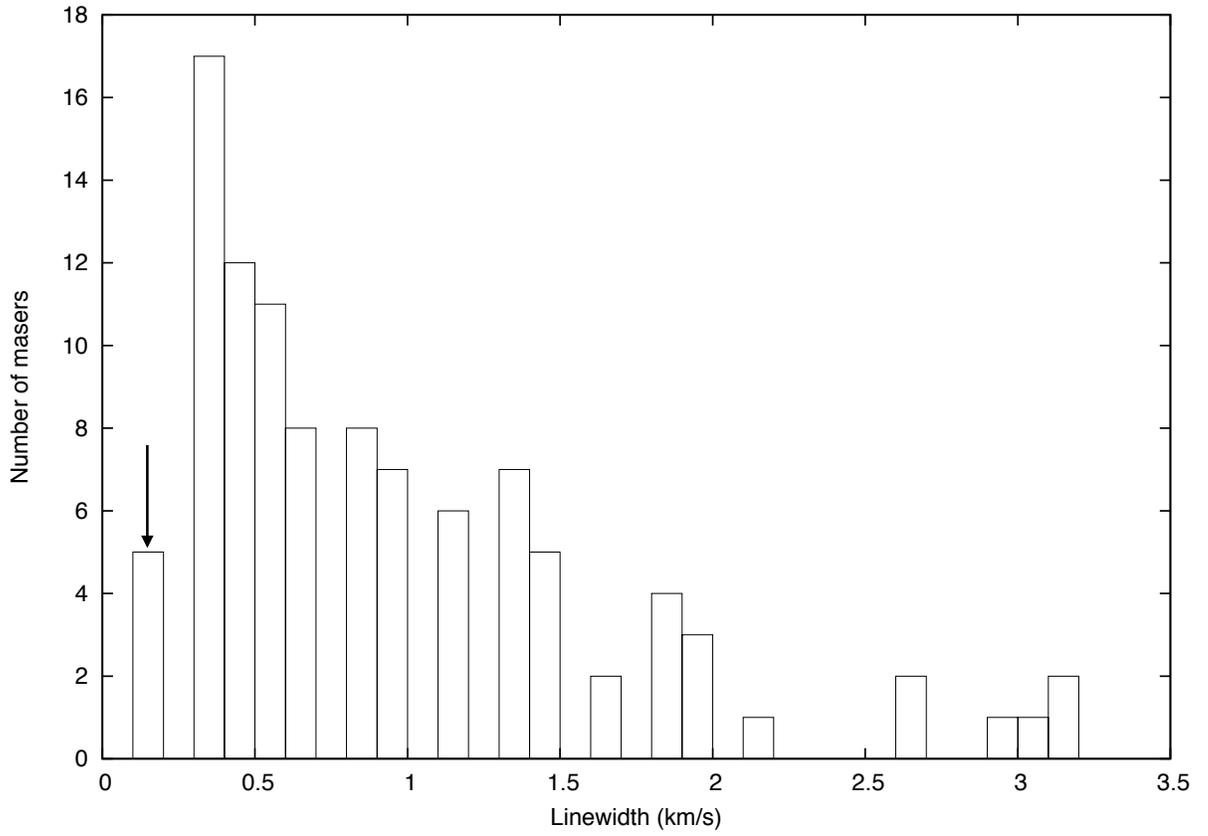}
\caption{Distribution of maser linewidths (FWZI) for the whole sample. The arrow indicates the bin corresponding to our spectral resolution.  
\label{FWZI}}
\end{center}
\end{figure}

Figure \ref{FWZI} shows the linewidth distribution (FWZI as reported in Table 2) 
of the masers we detected. The
peak of the distribution is at 0.3-0.4 km s$^{-1}$, and the median
value is 1.0 km s$^{-1}$. The linewidths range from $\geq$ 0.17 km
s$^{-1}$ (the channel width) to 3.15 km s$^{-1}$; we note that the
wider linewidths may correspond to multiple un-resolved maser
components and/or thermal contribution to the emission. Masers from
both groups, HMPO and UC HII regions, have velocities covering nearly
this full range; i.e., there is essentially no difference in the
linewidths between the two groups.

An extensive interferometric survey of southern class I methanol
  masers (at both 36 and 44 GHz) was recently published by Voronkov et
  al. (2014).  Their sample size (71 sources) was very similar to ours;
  their angular resolution ($6''$) was somewhat lower, but their spectral
  resolution (0.064 km s$^{-1}$) was nearly 3 times higher.  Overall, the
  two surveys are sufficiently similar that it is worthwhile to compare
  results.
  The linewidth distribution that we show in Figure 1 is generally
  similar to that found by Voronkov et al. (2014). Both distributions rise
  sharply from the spectral resolution of the observations, peak
  at about 0.3 km s$^{-1}$, and fall-off steadily at greater linewidths. 
  Our distribution is somewhat flatter (i.e., it falls off more slowly),
  which may occur because  Voronkov et al. (2014) use the effective FWHM, based
  on multi-component Gaussian fits,  while we use the FWZI.
  Voronkov et al. caution that all their narrow-line components arise
  from these multi-component fits, so that possibly the extreme narrowness
  of the distribution is a result of the fit systematics.  The fact that our
  distribution peaks at a similar narrow value suggests that these fit
  systematics probably do not substantially affect the results.

  Voronkov et al. (2014) report linewidths up to 8.0 km s$^{-1}$ and
  suggest that large linewidths ($>$ 1.5 km s$^{-1}$) seem to arise
  from weak wings of the maser emission, although they do not discard
  the possibility of quasi-thermal emission.  Our broadest linewidth
  is 3.2 km s$^{-1}$.  It is unclear if our narrower linewidths result
  from an intrinsic difference between the samples or if they are an
  observational artifact.  The VLA data presented here have higher
  spatial resolution and (on average a factor of two) better sensitivity than the ATCA data presented by Voronkov et al. (2014). It is possible that we spatially resolve different spectral
  components that ATCA sees blended in position.  Sensitive, high 
  resolution (both angular and spectral) observations of the broader
  sources from the Voronkov et al. (2014) should be conducted.

\subsubsection{Relative Velocity Distribution}

\begin{figure}
\includegraphics[scale=0.65,angle=270]{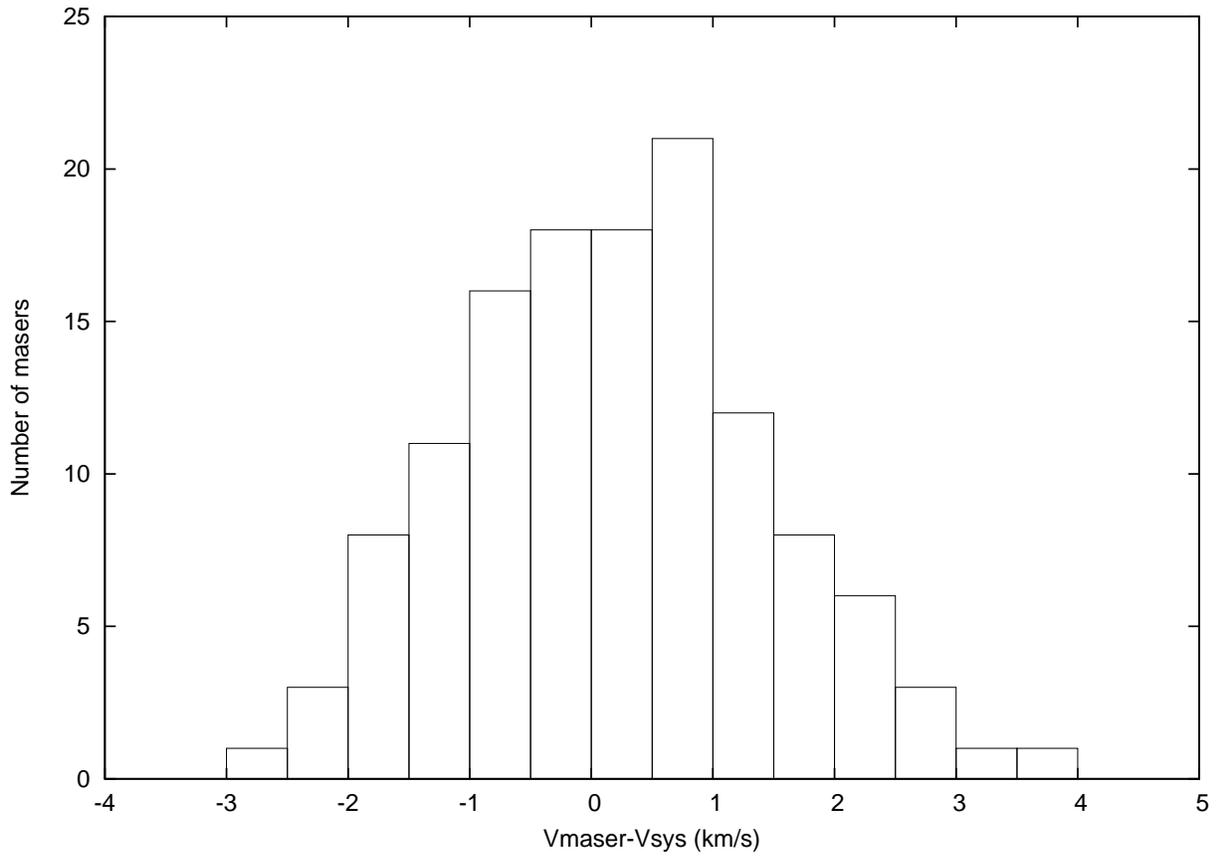}
\caption{Relative velocity distribution of the masers detected.   For the
systemic velocity we use the ammonia velocity reported by M96.
\label{relvel}}
\end{figure}

The relative velocity distribution of the methanol masers is shown in
Figure \ref{relvel}. We use the ammonia velocities reported by M96 for the
systemic velocity of the cloud, and plot the difference of the maser
velocities with respect to the cloud velocities. These relative
velocities range from $-$2.7 to $+$3.8 km s$^{-1}$. The distribution
peaks at $\sim$ 0.7 km s$^{-1}$, with a standard deviation of 1.2
km s$^{-1}$. The total range of velocities is less than 6.5 km
s$^{-1}$. This is substantially less than the 18 km s$^{-1}$ range
found for 6.7 GHz class II methanol masers by Szymczak \& Kus (2000),
but consistent with the 5 km s$^{-1}$ found by Fontani et al. (2010)
for class I methanol masers  (via single-dish observations). Voronkov et al. (2014) found that, apart from a small number of blue-shifted high-velocity masers (with offsets greater than 15 km s$^{-1}$), the distribution of relative maser velocities is centered around zero, with a mean value of $-0.57$ km s$^{-1}$, and standard deviation of $3.32$ km s$^{-1}$. Our distribution peaks at a marginally higher velocity and is noticeably narrower. A possible explanation for the latter difference is that we calculate the relative velocity with respect to ammonia, while Voronkov et al. (2014) use the middle of the 6.7 GHz methanol maser velocity range as the reference. We note that using ammonia as the reference, rather than 6.7 GHz methanol masers, results in relative velocities that differ by a few km~s$^{-1}$. For example, comparing the 44 GHz maser velocities of IRAS 20126+4104 reported in this paper to the 6.7 GHz maser velocities reported by Surcis et al. (2014) results in a relative velocity 2.7 km~s$^{-1}$ different from that found using ammonia as the reference.  The wider distribution reported by Voronkov et al. (2014) may be caused by this effect. However, the absence of high-velocity features with offsets greater than 15 km s$^{-1}$ in our data is due the limited bandwidth of our observations ($\sim$ 20 km s$^{-1}$).

Some authors argue that the small relative velocity of these masers
supports the idea that class I methanol masers arise in the interface
region between outflows and cores (see Plambeck \& Menten, 1990). We
discuss the relation of masers with molecular outflows in $\S$ 5.2 and 5.6.

\subsection{Comments on selected regions}

Because the sample extends well into the second galactic quadrant, many of our sources do not have survey data (IR or radio) that are readily available. Ancillary observations at arc-second resolution exist for some sources of our sample, including observations of molecular tracers that allow us to make a comparison with protostellar outflows. The source of information is diverse: millimeter, near- and mid-IR observations. Here we discuss these data in the context of our 44 GHz methanol maser detections. With the exception of Mol 45 (for which recently published IR data are now available) sources that are common to Paper I are not re-discussed here, since no significant differences were found by the new observations. Given the attention during the last years of the correlation between Spitzer/IRAC images and class I methanol masers (e.g., Cyganowski et al. 2009), we discuss that correlation separately ($\S$ 5.6).

\subsubsection{IRAS 05274+3345 (Mol 10)}

Mol 10 does not present a high level of star formation activity; its IRAS luminosity is 3 $\times$ 10$^3$ L$_{\odot}$, but molecular emission is very weak \cite[]{Estalella93}. About 30$''$ to the east of the IRAS source, centimeter and millimeter continuum emission reveal a cluster of five protostars (the AFGL 5142 region: Zhang et al. 2007; Hunter et al. 1999; Hunter et al. 1995). Associated with the continuum cores are at least three bipolar outflows. The main outflow (named outflow A by Zhang et al. 2007) was reported by Hunter et al. (1999) in a north-south direction in SiO emission, but recent mm/sub-mm observations show that at least two additional outflows are present in different directions (Zhang et al. 2007; see Fig. 3). Water masers and IR emission are coincident with shocked regions within the outflow lobes (Zhang et al. 2007; Hunter et al. 1995, 1999).

\begin{figure}
\begin{center}
\epsscale{1.0}
\plotone{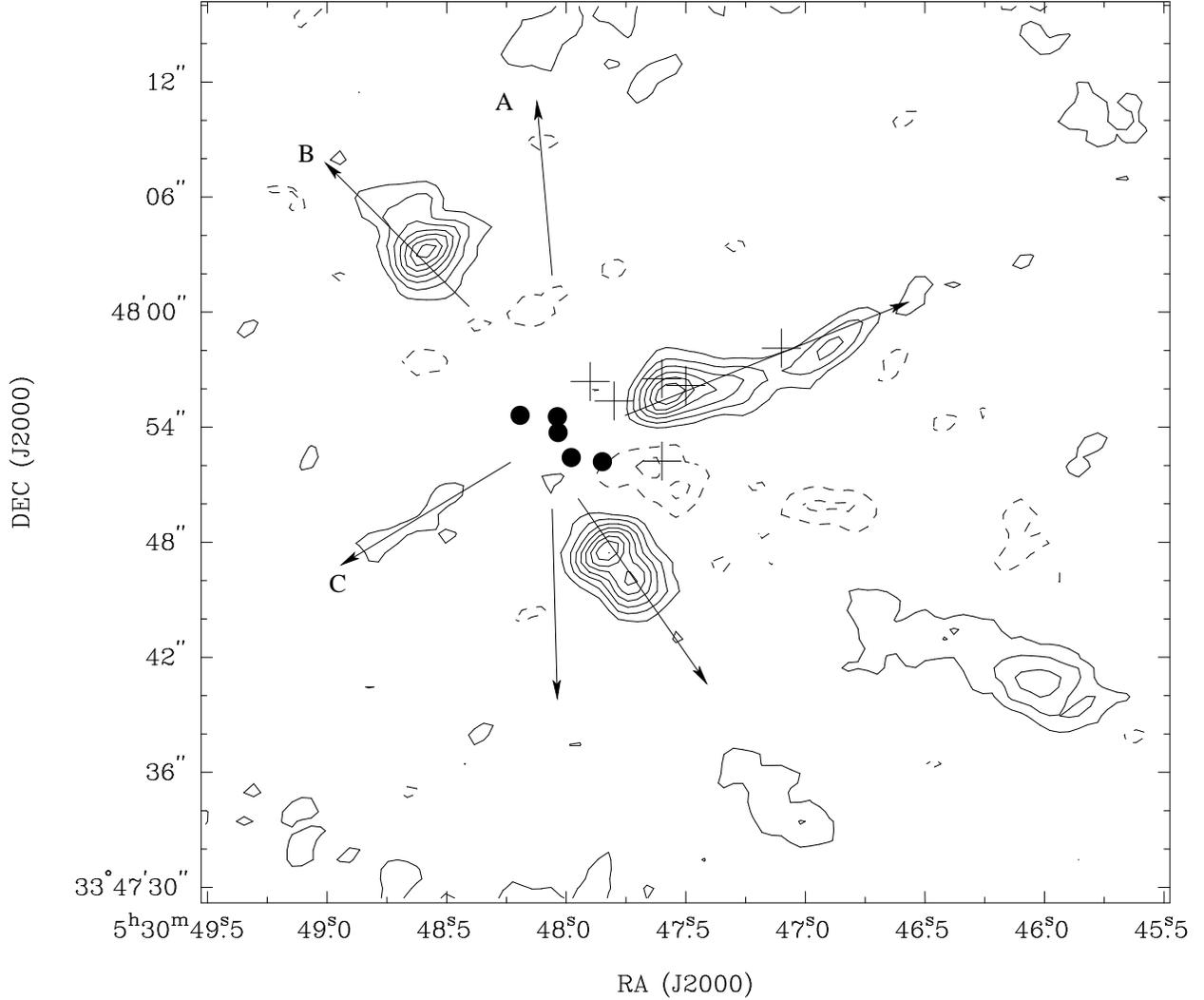}
\caption{Multiple outflows in the IRAS 05274$+$3345 region (Mol 10 or AFGL 5142). Contours show the integrated intensity CO (2--1) emission around the V$_{LSR}$ of $-16.4$ km s$^{-1}$ (width $\sim$ 10 km s$^{-1}$) from Zhang et al. (2007). This channel was selected in order better show the multiple outflows in the region (see Fig. 5 in Zhang et al. 2007). Filled circles denote millimeter continuum sources while plus symbols denote the six 44 GHz methanol masers we detect (the other two masers reported in Table 2 overlap with some of the six shown here). Arrows show the direction of the bipolar outflows labeled A, B and C, in the nomenclature of Zhang et al. (2007). Note that outflow A was seen in the SiO observations of Hunter et al. (1999), but not detected in CO.
\label{mol10}}
\end{center}
\end{figure}

Our VLA observations do not show maser emission near the IRAS source, but masers are found to the east, near the molecular outflows (see Fig. \ref{mol10}; also $\S$ 5.6). The positions of the maser components in AFGL 5142 confirm their relation with the outflow lobes, as shown in Figure 3. All maser components seem to be associated with the western lobe of outflow C and possibly the southern, inner part of outflow B. Despite the positional coincidence with the outflows, the maser velocities are within a few km s$^{-1}$ of the systemic velocity ($-$3.8 km s$^{-1}$), while Zhang et al. (2007) report terminal velocities of 25 km s$^{-1}$ for outflows B and C. Thus, if the masers participate in the dynamics of the outflows, they should be related to its low-velocity component or the swept-up material.

\subsubsection{IRAS 18144$-$1723 (Mol 45)}

\begin{figure}
\begin{center}
\epsscale{1.0}
\plotone{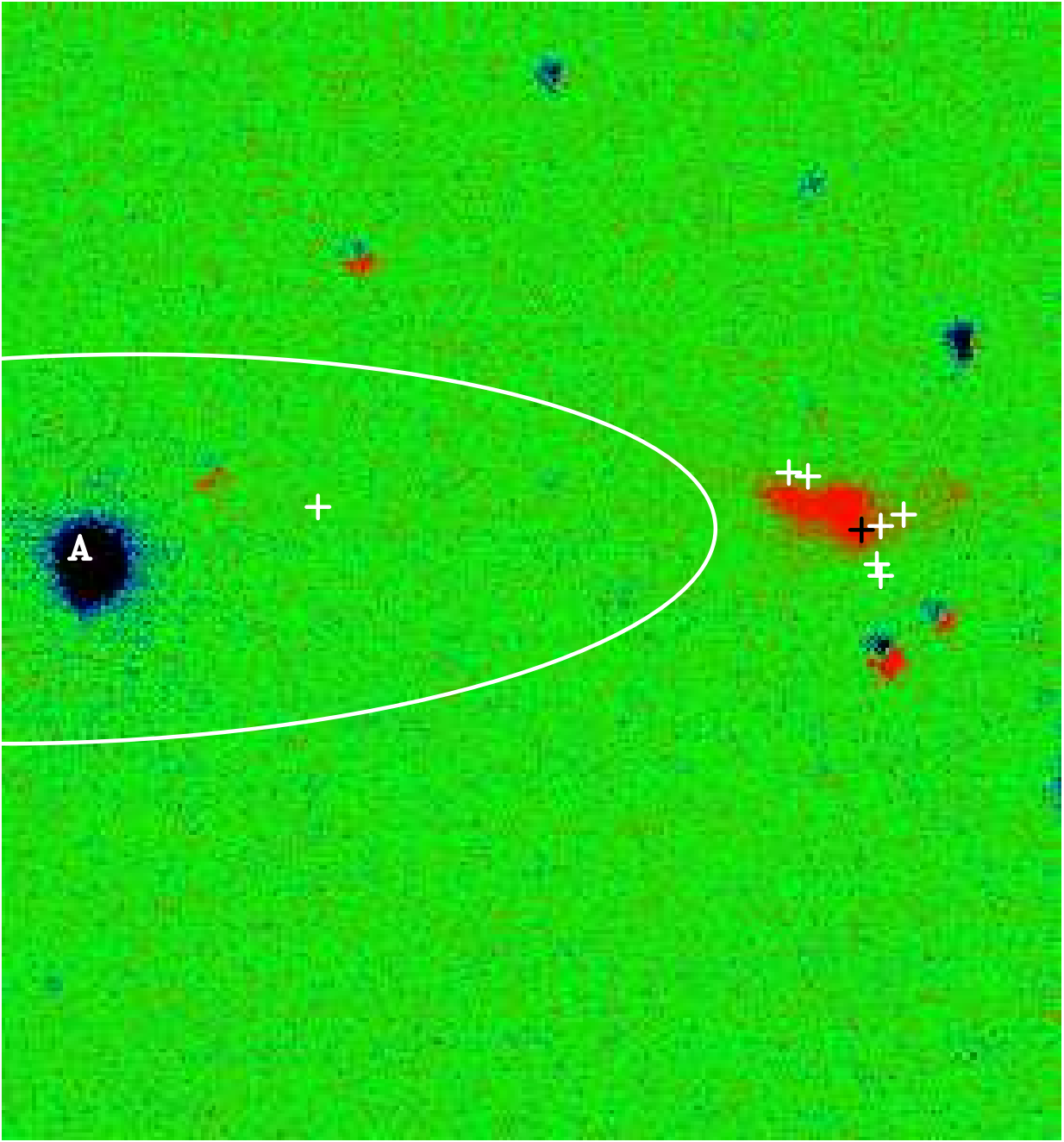}
\caption{IRAS 18144$-$1723 (Mol 45). In false calors we show the  continuum-subtracted H$_2$ image from Varricatt et al. (2010). Maser positions are shown by crosses. Nine of the maser spots we report are clustered near the bow-shock-like feature to the west of the image. The remaining two (overlapping) masers are further to the east, about 10$''$ from the infrared source `A'. The ellipse indicates the IRAS position.
\label{I18144}}
\end{center}
\end{figure}

Zhang et al. (2005) did not report an outflow in this field. Nevertheless, Varricatt et al. (2010) present an H$_2$ image showing a bowshock-like feature about 18$''$ to the west of the IRAS position (see Fig. \ref{I18144}). Coincident with the IRAS source, they detect a K-band continuum source which they call `A', that has some surrounding nebulosity.  They suggest that the bowshock-like feature originates from source A, and represents a highly collimated (factor 10) outflow at position angle  274$^{\circ}$.

We detect 11 maser spots in this field. Nine of these (two of them overlap) fall in a cluster within about 5 arcsec of the bowshock feature reported by Varricatt et al. (2010). The two remaining (overlapping) masers are located about 10 arcsec west of the infrared source `A' (Fig. \ref{I18144}).

\subsubsection{IRAS 18507+0121 (Mol 74)}

\begin{figure}
\epsscale{0.9}
\plotone{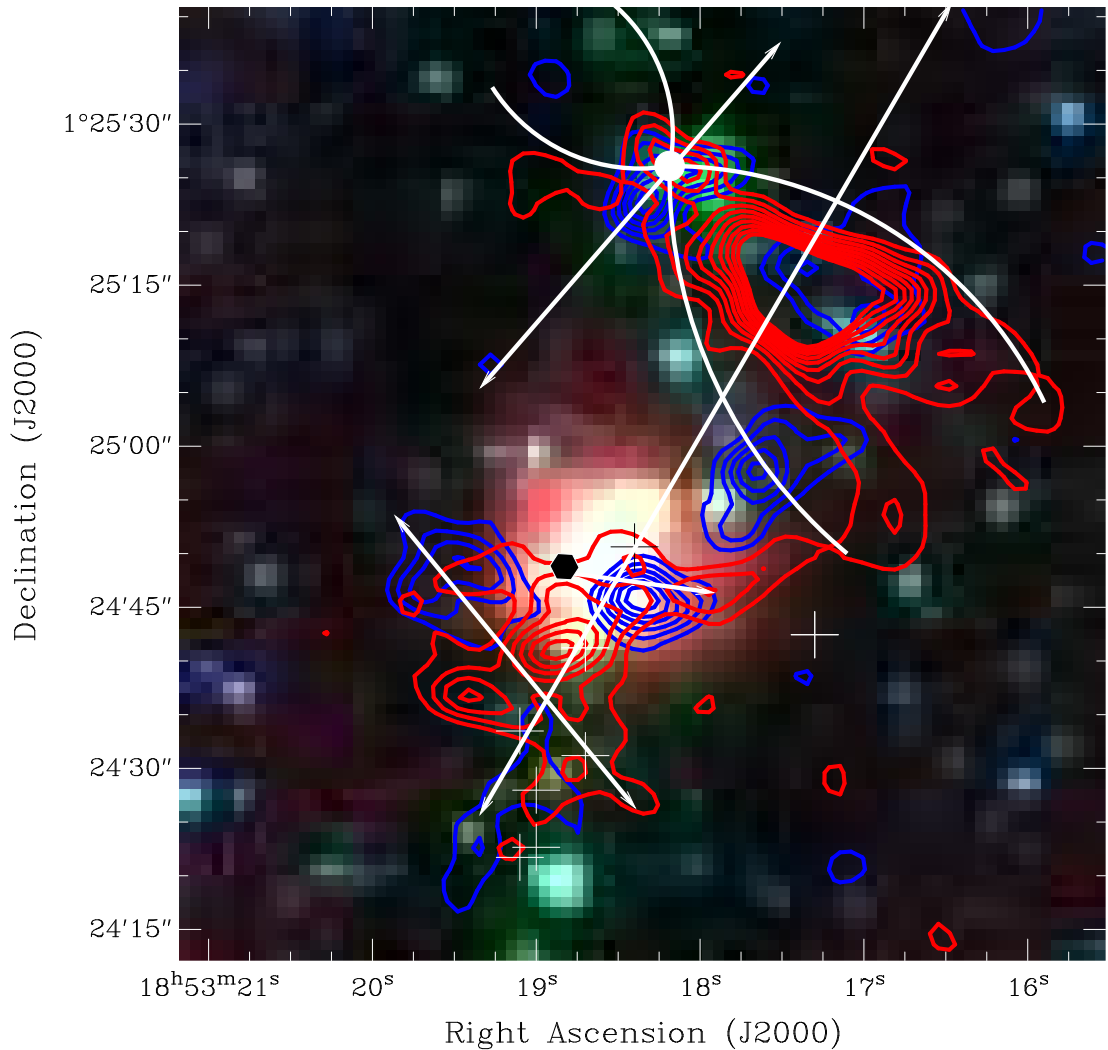}
\caption{
Overlay of high-velocity CO emission (red and blue contours) on the
three color image from Spitzer (3.6, 4.5, 8.0 $\,\mu$m) toward IRAS 18507+0121 (Mol 74; Shepherd et al. 2007). The eight plus symbols show the 44 GHz methanol maser positions from this paper, while the white circle and
black hexagon denote the millimeter source (MM1) and the UC HII region (MM2),
respectively, reported by Shepherd et al. (2007). The white arcs and lines show the outflow systems proposed by Shepherd et al. (2007). Note that the
maser spots are distributed closer to the UC HII region than to the
millimeter source, and seem to be associated with 4.5$\,\mu$m emission excess.
\label{I18507}}
\end{figure}

This massive star forming region is located in an infrared dark cloud
(IRDC) which hosts nine millimeter cores \cite[]{Shepherd07}. The
millimeter core MM2 is associated with the IRAS source and an UC HII
region \cite[]{Miralles94,Molinari98}; see Fig. \ref{I18507}. The centimeter
continuum flux is consistent with excitation by a ZAMS B0 star
\cite[]{Shepherd04}. VLA observations show H$_2$O masers toward MM1,
MM3 and MM4 \cite[]{Wang06}; only MM1 lies within the field shown in
Fig. \ref{I18507}. These cores have bolometric luminosities of 3.2 $\times
10^4$, 9.0 $\times 10^3$, and 1.2 $\times 10^4$ L$_{\odot}$,
respectively \cite[]{Rathborne05}, suggesting that they are high-mass
protostars.

Shepherd et al. (2007) report three outflows centered on or near the
UC H\,\small II region (G34.4+0.23; i.e., MM2), and two other outflows from the
millimeter core G34.4+0.23 MM (MM1). Fig. \ref{I18507} shows high-velocity CO
emission tracing the molecular outflows overlaid with the 44 GHz maser
positions. All the maser spots are located around the UC H\,\small II region,
and some of them are coincident with the outflow lobes. No maser
emission was found toward the outflows from MM1, although the VLA
primary beam covered that region. This information suggests that, as in the case of Mol 10 (AFGL5142), class I methanol masers are located toward the more-evolved objects of the cluster
(i.e. outflows from sources that already developed an UC H\,\small II region).

\subsubsection{IRAS 18517+0437 (Mol 76)}

Varricatt et al. (2010) report an  embedded source located 
about 12$''$ south-west of the IRAS position. 
No collimated structure was found in H$_2$; rather, the  H$_2$ line
emission appears as a faint, arc-like feature $\sim 5''$ north-east of
the embedded source.   
We detect three maser spots, distributed around the arc-like feature. 

Interferometric observations
by Schnee \& Carpenter (2009) detected two masers in the 95
GHz class I transition.  These masers (which they label $a$ and $b$)
correspond (in position and velocity) to our masers 3 and 2, respectively.

\subsubsection{IRAS 19374+2352 (Mol 109)}

\begin{figure}
\begin{center}
\epsscale{1.0}
\includegraphics[angle=-90,scale=0.8]{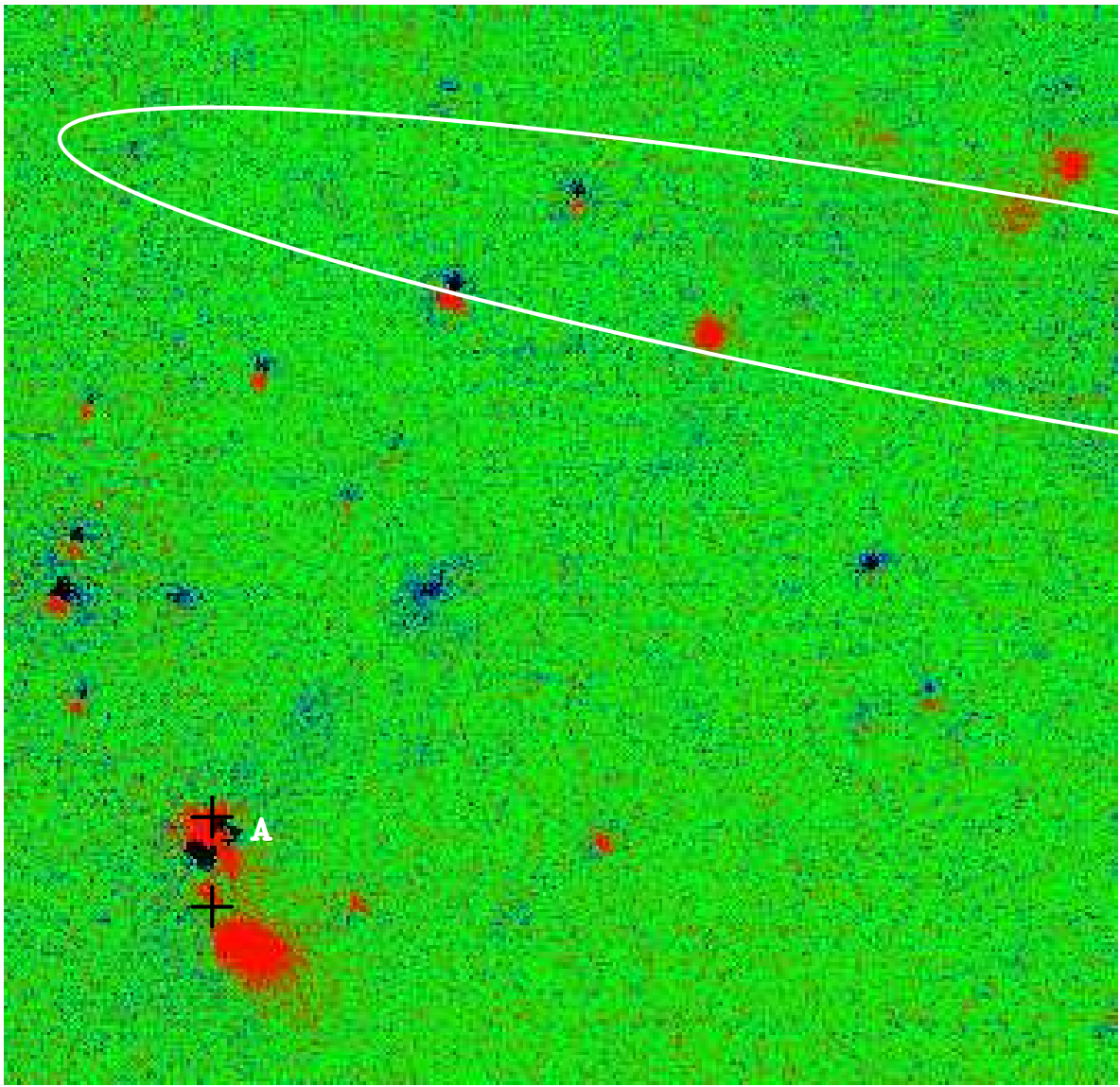}
\caption{Same as Fig. \ref{I18144} but for IRAS 19374$+$2352 (Mol 109). A few arcsec to the south of source 'A', H$_2$ emission is seen, possibly tracing a collimated outflow. The 44 GHz masers are located within 5$''$ of source 'A'.
\label{I19374}}
\end{center}
\end{figure}

Varricatt et al. (2010) find H$_2$ emission associated with
this source (which they label `A') and note that it corresponds to a
pair of stars separated by 1\rlap.{$''$}5 (see Fig. \ref{I19374}). A few arcsec to the south
of source 'A', H$_2$ emission {\it is} seen, possibly tracing a
collimated outflow. Based
on their data and previous mm-wave observations, Varricatt et
al. (2010) infer the presence of multiple molecular outflows in the
region, although they cannot identify the driving sources. If one of
these stars is driving a molecular outflow, then the methanol masers
would evidently arise near the base of the outflow, rather than at the
interaction region of the outflow lobes with the interstellar medium. 
As seen in Fig. \ref{I19374}, the 44 GHz masers are located within 5$''$ of source `A'. 

\subsubsection{IRAS 20050+2720 (Mol 114)}

\begin{figure}
\begin{center}
\epsscale{1.0}
\plotone{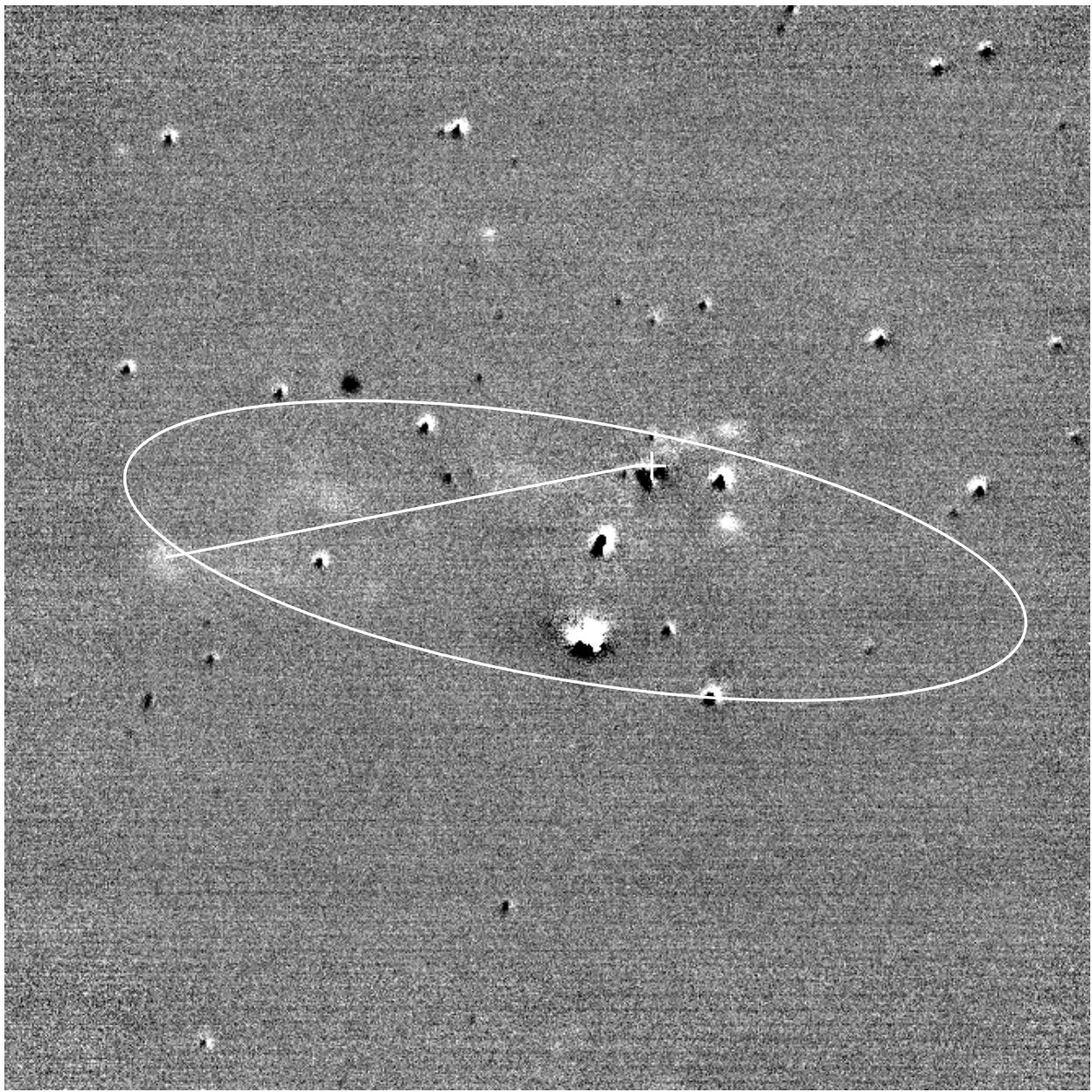}
\caption{Same as Fig. \ref{I18144} but for IRAS 20050$+$2720 (Mol 114). Varricat et al. (2010) suggested several outflows in the region. One lobe of the main outflow of the region is indicated by the white line, at the base of which, projected near to the proposed driving source ($<$ 2$''$), we see the 44 GHz methanol maser.
\label{I20050}}
\end{center}
\end{figure}

We detect a single maser component in this field, coincident
with the millimeter continuum source OVRO 1 reported by
Beltr\'an et al.  (2008).  Beltr\'an et al. conclude that 
two molecular outflows are present, and that their driving sources are
OVRO 1 and a YSO within the 8000 AU (11\rlap.{$''$}3) envelope surrounding 
OVRO 1. The maser position that we report is less than 1$''$ from the
OVRO 1 position, suggesting that the maser is associated with the east-west
outflow, driven by OVRO 1, and that it arises near the base of the outflow.
We caution that this interpretation is not unambiguous.  Beltr\'an et al.
suggest that the two outflows interact with one another; it is possible that
the maser arises in an interaction region, rather than at the driving
source of one of the outflows.  Moreover, Varricatt et al. (2010) suggest 
that additional outflows might be present.  If so, and if one of these were
oriented along the line-of-sight, the maser could arise rather far from
the outflow origin, but appear projected near ($<$ 2$''$) to the driving source (see Fig. \ref{I20050}).

\subsubsection{IRAS 20056+3350 (Mol 115)}

Although M98 did not detect radio continuum emission in this field, Jenness et
al. (1995) report weak (0.6 -- 0.9 mJy) 8 GHz emission located about
2$''$ from the maser positions.  Although this continuum emission ---
located so close to the maser positions --- is interesting, we do not
count Mol 115 as a source with a UC HII region, because the quite low
flux density suggests that the origin of the emission is not 
photoionization by a young, early-type star, but rather is due to some
other mechanism (e.g., Rodr\'\i guez et al. 2012).

\subsubsection{IRAS 20062+3550 (Mol 116)}

\begin{figure}
\begin{center}
\epsscale{1.0}
\plotone{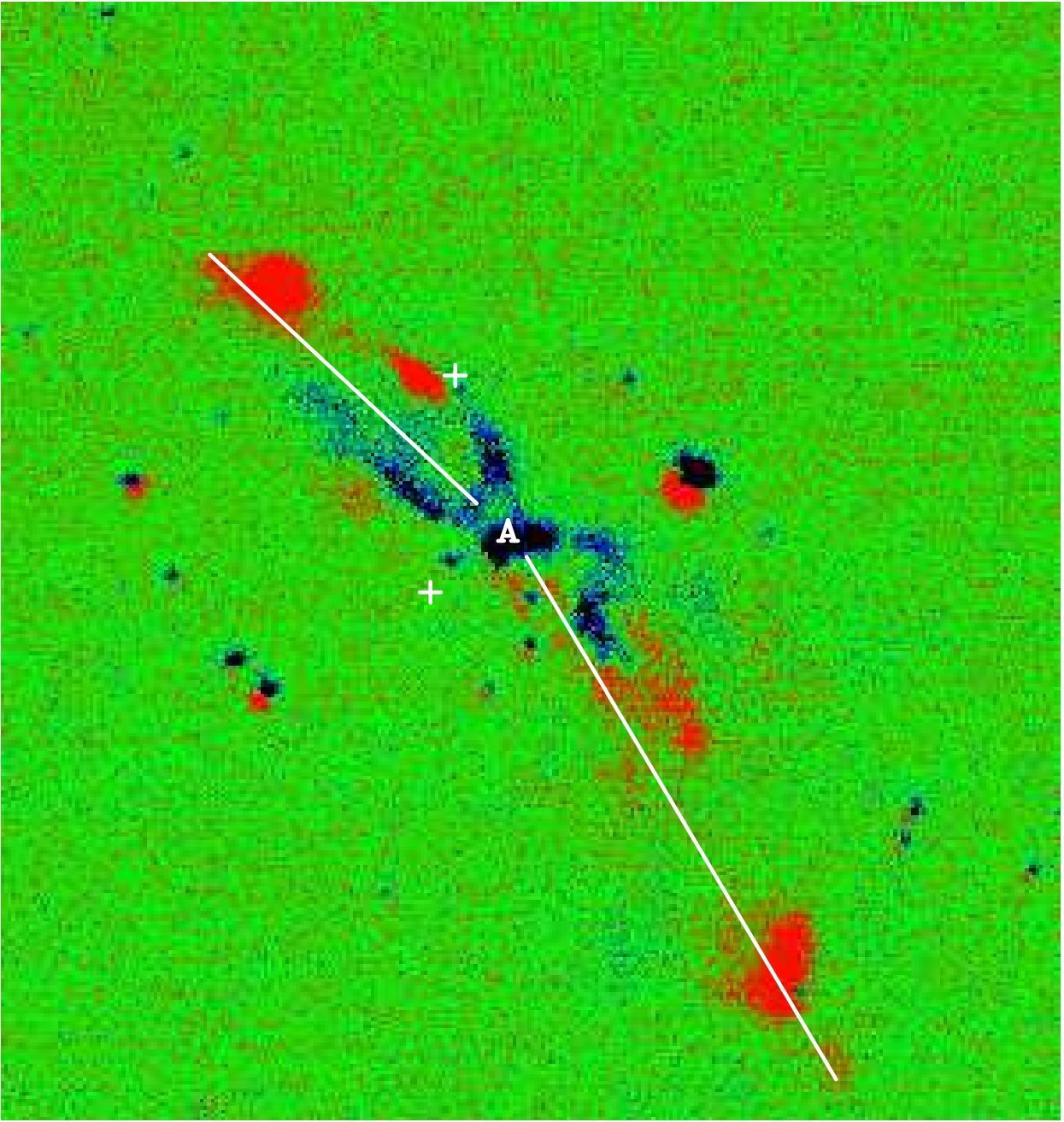}
\caption{Same as Fig. \ref{I18144} but for IRAS 20062$+$3550 (Mol 116). The proposed source 'A' that drives a bipolar outflow, which has a lobe extending 10$''$ to the north-east and another extending 16$''$ to the south-west (indicated by white lines). Two 44 GHz detected relatively close (1\rlap.{$''$}4 and 4\rlap.{$''$}9) to the source 'A'.
\label{I20062}}
\end{center}
\end{figure}

Mol 116 hosts one of the larger molecular outflows reported by
Zhang et al. (2005). The outflow has a far more spectacular appearance
in the K-band and H$_2$ images of Varricatt et al. (2010), as seen in Fig. \ref{I20062}. 
They report an IR source `A' as the probable driving source of the outflow, which
has a lobe extending 10$''$ to the north-east and another extending $16''$ to 
the south-west.  The two masers we detect are located relatively close
(1\rlap.{$''$}4 and 4\rlap.{$''$}9) to source A.  This is in contrast with 
the case of IRAS 20126$+$4104 (see Paper I) in which the masers coincide with the tips of
the outflow lobes.  In Mol 116 the masers seem to be located in the interaction 
zone between the inner part of the cavity wall and the surrounding medium. 
This behavior may also occur in Mol 114 and, possibly, Mol 109. Alternatively, the emission could arise from an undetected pole-on outflow produced by another central source. 

\subsubsection{IRAS 22506+5944 (Mol 151)}

Single-dish and interferometric CO observations of the outflow in Mol
151 were reported by Wu et al. (2005) and Su et al. (2004),
respectively.  The single-dish map shows a north-south outflow, while
the interferometer map (including single-dish data for the zero
spacing) shows an east-west outflow.  In both cases, the total
angular extent of the molecular emission is of order 1 arcminute. The
driving source of the outflow(s) has been attributed to both IRAS
22506+5944 and to the near-IR source S4 (2MASS J22523871+6000445; e.g.,
Xu \& Wang, 2010).

The S4 position is nominally outside the IRAS error ellipse for
22506+5944.  Nevertheless, we note that the 2MASS image shows a
compact cluster of five sources that {\it IRAS} could not have
resolved.  We consider the IRAS source to correspond to the cluster,
not to its individual members, and disregard the precise IRAS
coordinates.

More importantly, we note that all six maser components that we detect
are within 9$''$ of the S4 position (see Figure 7 of Su et al.
2004) while the two outflow lobes span 1$'$ in the east-west direction.
Thus, as in several previous cases, the masers might trace the base of
the outflow and not the terminus of the outflow lobes.

\subsection{44 GHz methanol masers in HMPO's and UC HII regions}

Molinari et al. (1998) observed 67 sources from their larger sample in the 6 cm continuum with the VLA, and  detected emission at the 1 mJy level (or higher) in 37 (55\%) of the sources.  If the centimeter continuum emission was within 40$''$ of the IRAS position, then they considered the continuum emission to be an IRAS-related UC H\,\small II region; 22 sources were in this category. In Table 1 we list sources as ``UC H\,\small II regions'' if the free-free emission is within 40$''$ of the corresponding IRAS source. As we note in \S2, in some cases this centimeter continuum may arise from some other class of object than a UC H\,\small II region (e.g., a thermal jet).  On the other hand, if there is no detected centimeter free-free emission ($\gtrsim$ 1 mJy), or if it is more than 40$''$ distant, we define the source as a ``HMPO''.  

\begin{figure}
\includegraphics[angle=270,scale=0.7]{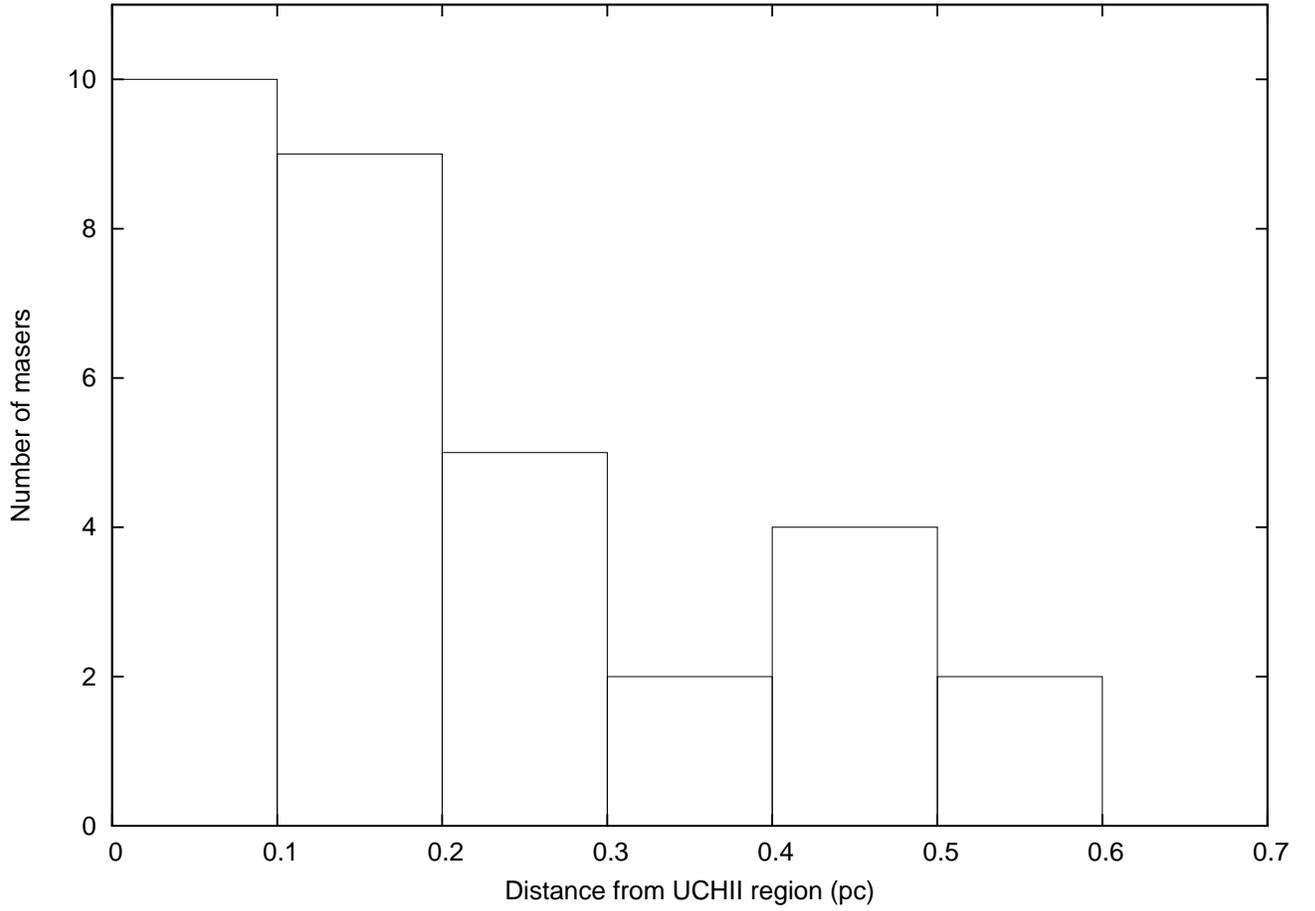}
\caption{Histogram of the projected distance of maser components from UC HII regions. 
\label{maser-UC}}
\end{figure}

We show the location of the masers with respect to the continuum emission (from M98) in Fig. \ref{3col}, where the centimeter continuum emission from Molinari et al. (1998) is shown by contours. For the 13 UC HII sources showing maser emission, we measure the projected distance from each maser component to the peak position of the continuum emission (based on the source distances in Table 1 and the continuum positions reported by M98). The resulting projected distances are shown in Fig. \ref{maser-UC}. The shortest separation measured is 0.005 pc and the largest is 0.6 pc. In four sources some maser components are projected against the centimeter emission. This result is consistent with that of Kurtz at al. (2004) that class I methanol masers can be found in the immediate vicinity of UC HII regions.

Excluding the three sources that we were unable to calibrate, we find
that 13 of 24 ($54$ \%) of the
UC H\,\small II fields show maser   
emission, while 16 of 42 HMPO fields (38 \%) show masers. The sample
sizes are not large enough for this result to have a firm statistical
basis, nevertheless, these detection rates suggest that class I
methanol masers are more common in the presumably more-evolved regions
of the Molinari sample that present centimeter free-free emission. Voronkov et al. (2014) find a much lower association of class I maser emission with centimeter continuum emission (17 of 71 sources or 24\%). However, as they note, their continuum data are very inhomogeneous, so this association rate should be treated as a lower limit. We caution that assigning evolutionary stages based on the presence or absence of particular maser species may not be always reliable. We refer the interested reader to \S4.5 of Voronkov et al. (2014), which presents an enlightening discussion of this point.

\subsection{Color-color analysis}

\begin{figure}
\begin{center}
\includegraphics[angle=270,scale=0.8]{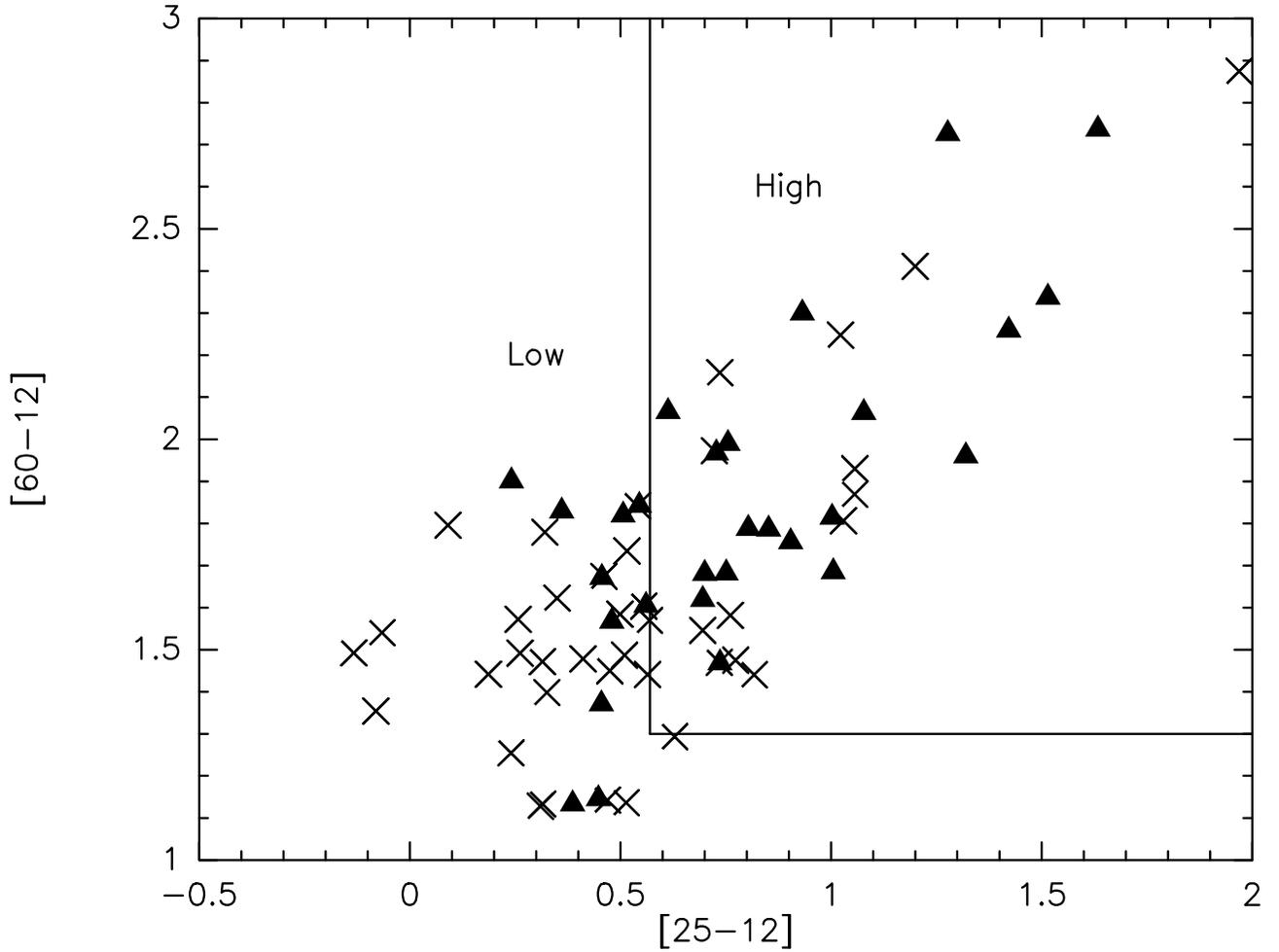}
\caption{[25-12] vs. [60-12] color-color plane. Maser detections are indicated by triangles while non-detections are shown by crosses. The boxed region in the upper right indicates the portion of the color-color plane satisfying the Wood \& Churchwell (1989) criteria for colors typical of UC HII regions. Following the classification of M96, sources in this region are denoted $High$ and sources outside are denoted $Low$.
\label{irascol}}
\end{center}
\end{figure}

Color-color diagrams applied to different maser species have proven
useful to search for trends in maser and far-IR properties (e.g.,
H$_2$O: Palla et al., 1991; class II CH$_3$OH: Szymczak \& Kus, 2000;
OH: Edris et al. 2007). In Figure \ref{irascol} we present the color-color diagram
[25-12] versus [60-12]; the limits for {\it High} and {\it Low}
sources follow the M96 classification. Excluding  the three sources we
were unable to calibrate, our sample contains 30 {\it High} and 36 {\it
  Low} sources. The maser detection rates are 19 out of 30 for {\it High}
 ($\sim$ 63 \%) and 10 out of 36 for {\it Low} ($\sim$ 27 \%).
 As pointed out previously, the {\it High} classification
corresponds to sources with colors meeting the Wood \& Churchwell
(1989) criteria for UC H\,\small II regions. Inspection of the
color-color plot shows that in general the detection rate increases
toward the redder corner (upper right) of the box. In particular, a
maser detection rate of $70 \%$ is found in the region [25-12] $>$ 0.90
and [60-12] $>$ 1.99. This is in agreement with Fontani
et al. (2010), who found a 3 times higher detection rate in the {\it
  High} region compared to the {\it Low} region.

Palla et al. (1991), using a much larger sample, found that water
masers are common in sources with the Wood \& Churchwell (1989)
colors. Their water maser detections were bounded by [25-12] $\geq$
0.1 and [60-12] $\geq$ 1.1. They found that setting the [25-12] index
to $\geq$ 0.50 did not affect the detection rate, and concluded that a
[25-12] index larger than 0.5 is a reliable indicator for far-IR sources
associated with H$_2$O maser emission. They found a $50 \%$ detection
rate for H$_2$O masers when [60-12] $> 1.8$ and [25-12] $> 0.9$. For OH
masers, Edris et al. (2007) combined the M96 and Sridharan et
al. (2002) samples, and report masers for color indices [25-12] $\geq$ 0.23 and [60-12] $\geq$ 1.1 (i.e., higher detection rates toward redder
sources). Also, sources with OH masers dominate over non-maser sources
for regions with [25-12] $>$ 1.2 and [60-12] $>$ 2.2. In the case of
class II CH$_3$OH masers, Szymczak \& Kus (2000) found that the highest
detection rates were for intermediate colors. Their maximum detection
rate was found near 0.8 $<$ [25-12] $<$ 1.0 for [60-25] $<$ 0.8.

Our results show that class I CH$_3$OH masers follow similar trends
to OH, H$_2$O and class II CH$_3$OH masers with respect to the 
far-IR colors of the host region.  
We caution that any color-color trend between {\it High} and
{\it Low} sources must be tested {\it independently} from a luminosity
effect.  M96 report a one-quarter dex higher luminosity for the {\it High}
sources compared to the {\it Low} sources (see their Table 4 and Figure 6).
Thus, differences between the two groups may arise from differing 
luminosities, rather than differing colors. 
  
\subsection{Maser luminosity and IRAS properties}

To discern the relation of maser emission with other star formation
processes, we looked for correlations between the isotropic maser
luminosity and the parameters of the central source. 
The maser flux density of each field and the
source distance (taken from M96) were used to calculate the
isotropic maser luminosity.  In the case of multiple maser components,
we summed the individual flux densities and used this value to obtain
the luminosity.

\begin{figure}
\includegraphics[scale=0.8,angle=270]{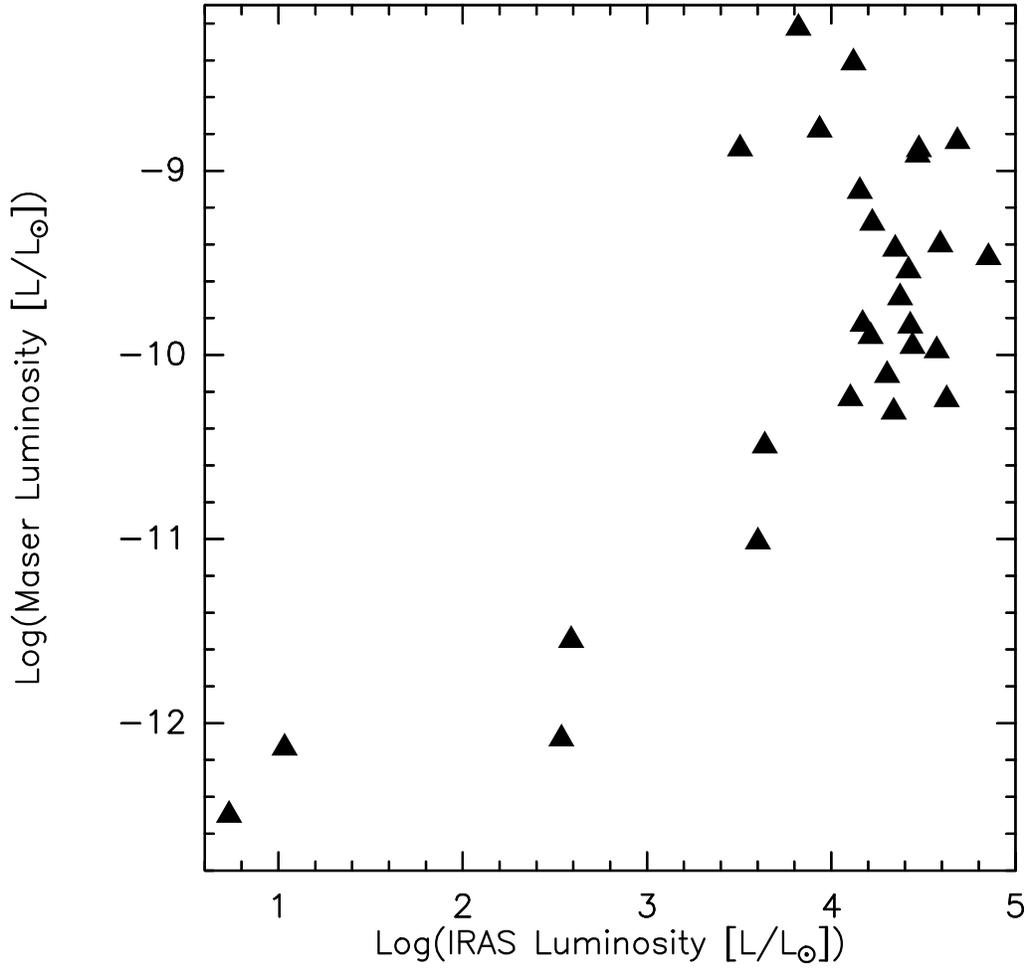}
\caption{Plot of isotropic maser luminosity versus IRAS luminosity.
\label{lumirel}}
\end{figure}

Figure \ref{lumirel} shows a plot of isotropic maser luminosity versus IRAS luminosity. For IRAS luminosities greater than $10^3$ L$_\odot$ there is no obvious correlation between the maser and infrared luminosities. The four sources with IRAS luminosities below $10^3$ L$_\odot$ have maser luminosities about two orders of magnitude lower than the rest of the sample, suggesting that there is a tendency toward higher maser luminosity as the IRAS luminosity increases. However, it is possible that the kinematic distances reported by M96 are inaccurate. In particular, we note that the distances of these four low-luminosity sources are uncertain and range from 80 to 730 pc, thus, their luminosity could have been understimated. If these distances are correct, then the sources Mol 38, Mol 114, Mol 116, and Mol 121 are low-to-intermediate mass star-forming regions, {\it not} high-mass star-forming regions. If so, we detected methanol masers from non-high-mass star-forming regions --- unusual for methanol masers (Kalenskii et al. 2013).

In summary, we consider the distances, luminosities, and classifications of these four sources to be uncertain.  A definitive determination of their distances is needed before we can ascertain the significance of the trend of lower maser luminosity with lower IRAS luminosity, and to investigate whether new ``low-mass'' methanol masers have been found.

\subsection{Relation with mid-IR emission from Spitzer/IRAC}

A relation between class I methanol masers and molecular outflows was
proposed by Plambeck \& Menten (1990) based on interferometric observations of
the 95 GHz transition in DR21. They found that the maser positions
coincided with H$_2$ emission from shocked gas along the outflow
lobes. Subsequent interferometric observations have supported the
outflow-maser association
\cite[]{Johnston92,Kurtz04,Voronkov06,Araya09}. Several of the sources discussed
in section 5.2 exemplify this relation. To study the maser--outflow 
connection in our sample, we use publicly available Spitzer images.

An important indicator of shocked regions in outflows is the mid-IR
emission from ro-vibrational H$_2$ transitions and/or CO bandheads
covered by the Infrared Array Camera (IRAC; Fazio et al. 2004). The
$\lesssim$ 2$''$ angular resolution of IRAC is similar to that of many
millimeter interferometric observations of CO and SiO lines, allowing
for useful comparisons. The 4.5$\,\mu$m band includes emission from
HI Br$\alpha$, CO(J=1-0) P(8) fundamental bandhead, several H$_2$
lines, and some forbidden molecular lines; notably, no PAH emission is
found in this band. Thus, the 4.5$,\mu$m band mostly traces shocked
and/or ionized gas (Shepherd et al. 2007). Very bright H$_2$ emission
and/or broad CO bandhead emission can produce an excess in the
4.5$\,\mu$m band, with respect to the 3.6$\,\mu$m and 8.0$\,\mu$m
bands (e.g., Shepherd et al. 2007, Smith et al. 2006). A 4.5$\,\mu$m
excess therefore indicates the presence of shocked molecular gas. We
searched the Spitzer archive for IRAC observations toward our sample
of massive protostars. Not all the fields were observed by Spitzer. Most of the images retrieved were observed as
part of the GLIMPSE survey (Benjamin et al. 2003).

In Fig. \ref{3col} we show the three-color images of the 3.6, 4.5, and
8.0$\,\mu$m bands, with overlays of the 44 GHz maser spots. A spatial
coincidence of the masers and the mid-IR emission is seen in most of
the sources. In at least half of the sources we find a 4.5$\,\mu$m
excess, indicating the presence of shocked gas, possibly related with
outflows.  In most cases, the methanol masers lie close to or coincident
with the 4.5$\,\mu$m excess.   

A similar comparison between 44 GHz class I methanol masers and
4.5$\,\mu$m excess emission is given by Cyganowski et al. (2009). They
observed 44 GHz class I masers with the VLA toward a GLIMPSE-selected
sample of sources with extended 4.5$\,\mu$m  excess (the
so-called Extended Green Objects or EGOs). Their observations show
several cases in which class I methanol masers surround the EGOs. This pattern is also found in some of our observations; see, for example, IRAS 18024$-$2119 (Mol 38) and IRAS 20126+4104 (Mol 119) shown in Figure 12. This behavior suggests that class I methanol masers are
tracing the interface between the outflow lobes and the surrounding
material, as originally proposed by Plambeck \& Menten (1990). In addition,
the small relative velocities reported in \S4.2.2 may relate these
masers with the post-shock gas (entrained material) in outflows.

This spatial coincidence in a substantial fraction of our sample is
strong circumstantial evidence to support the idea of a physical 
origin of the masers in shocked, possibly outflowing gas.
Nevertheless, we note that about half of the Spitzer images show
{\it no} evidence for a  4.5$\,\mu$m excess, despite the fact that 44 GHz
masers {\it are} detected.  Although the maser emission usually is closely
associated with mid-IR emission, there are at least two cases (IRAS
18144$-$1723 and IRAS 18565+0349) in which the masers appear to be
isolated, with no nearby mid-IR emission. 
A plausible explanation for these two cases is that the masers trace shocks 
from a very deeply embedded object, similar to the case of DR21(OH) in which the 
outflow and driving source are not detected in the mid-IR (Araya et al. 2009).

Particularly intriguing is our finding that in four cases, the
masers may be located close to the putative base of a molecular
outflow.  As we mention in \S4.3, the projected maser positions of
various sources fall close to the presumed driving source of the
outflow.  In particular, the Molinari sources 109, 114, 116, and 151
show this behavior.  Of course, lying close to the base of the outflow
does not preclude the possibility that the masers arise in shocked gas, and perhaps this is produced by an orientation affect [e.g., a pole-on (undetected) outflow].
The interpretation of this result is not yet clear; but it is in contrast
to the finding of other studies (e.g., Kurtz et al. 2004, Araya et al. 2009,
 Cyganowski et al. 2009)  in which the masers are found at the outer 
end of the outflow lobes, rather distant from the driving source.

In \S5.2 (see also Figures 3 and 5) we showed two cases in which class
I masers in regions with multiple outflows appeared toward
the more-evolved sources. In these two cases the outflows where the 
masers were found are also related to HII regions, suggesting that class I
masers may trace an evolved phase of the outflow. 
Other authors have reported the relation of class I methanol masers and HII regions, 
suggesting that expanding HII regions might be another shock environment in which class I 
masers are excited (e.g., Voronkov et al. 2010, Sobolev 1992).

In a larger context, this and other surveys of class I methanol masers not only suggest that these masers may trace shocks at different stages, but may even trace shocks arising from a number of different phenomena occurring in star-forming regions. Class I masers in regions with little or no mid-IR emission may trace deeply embedded shocks from 'young' bipolar outflows or cloud-cloud collisions (this work, Araya et al. 2009, Plambeck \& Menten, 1990). In IR-bright regions with a 4.5 micron excess, and therefore with strong shocks, the class I masers are found tracing bow-shocks of 'young' outflows (e.g., this work, Cyganowski et al. 2009). For regions with cm-continuum emission, the class I masers may trace bow-shocks from 'old' outflows and/or expanding HII regions (e.g., this work, Voronkov et al. 2010, Voronkov et al. 2014). 

\section{Summary and Conclusions}

The principal results of this survey of 44 GHz class I methanol masers are the following:

A 38\% (16/42) maser detection rate was found toward HMPO candidates, while 54\% (13/24) was the detection rate in fields with candidate UC HII regions. These detection rates suggest that although class I methanol masers are found in both HMPO and UC HII regions, they are more common in more-evolved regions.  

Toward UC HII regions, the projected distance from the maser spots to the peak of the continuum emission was between 0.01 and 0.6 pc. In some cases the masers are projected onto the ionized regions.

In comparison with other similar interterferometric data sets, our observations show narrower linewidths, likely due to our higher spatial resolution. Also, the relative velocity distribution is noticeably narrower, but in this case possibly due to the different method used to determine the cloud velocity. High-velocity class I methanol masers reported by other authors may have been missed by our observations, because our limited bandwidth. 

In two fields, possessing both young and old outflows, the masers were only found toward the more-evolved outflows. This is suggestive that class I methanol masers also trace later stages of outflow evolution. However, other authors have reported these masers towards tracers of 'young' outflows. Therefore, assigning evolutionary stages based on the presence or absence of particular maser species may not be always reliable.

Based on a comparison with observations of molecular outflows in the millimeter and IR, we find evidence that the masers are sometimes located at the interface between the molecular outflow and the surrounding material. In addition, the small relative velocities ($\leq$4 km s$^{-1}$) between the masers and the ambient molecular material suggests that the former are located in post-shock regions (entrained material) of molecular outflows. The presence of a number of class I masers that are {\it not} obviously associated with shocked gas, and which in some cases appear to arise very close to the base of the outflow, suggests that more than one scenario may be needed to understand the occurrence of 44 GHz class I methanol masers.

This and other surveys of class I methanol masers not only suggest that these masers may trace shocks at different stages, but may even trace shocks arising from a number of different phenomena occurring in star-forming regions: young/old outflows, cloud-cloud collisions, expanding H\,\small II regions, among others.

\acknowledgments

We thank the anonymous referee for the thoughtful reading and detailed comments that helped to improve the quality of this paper. AGR is grateful to CRyA-UNAM for its support to this project which formed part of his masters thesis at that institution. AGR is
supported by Consejo Nacional de Ciencia y Tecnolog\'ia, through the
program C\'atedras CONACYT para J\'ovenes Investigadores. We thank
D. Shepherd and K. Qiu for providing us with the CO data for I18507
and AFGL5142, respectively. PH acknowledges partial support from NSF
grant AST-0908901. SK acknowledges partial support from UNAM DGAPA
grant IN 114514. L.L. acknowledge the support of DGAPA, UNAM, CONACyT, Mexico, and the von Humboldt Foundation for financial support.

{\it Facilities:} \facility{VLA}.

\clearpage
\begin{deluxetable}{rccccccccr}
\tabletypesize{\scriptsize}
\tablecaption{Observed Source List}\label{tbl-1}
\tablewidth{0pt}
\tablehead{
\colhead{Mol} & \colhead{IRAS} & \colhead{$\alpha$(J2000)} & \colhead{$\delta$(J2000)} & \colhead{Central Vel} & \colhead{Dist. \tablenotemark{b}} & \colhead{Synth Beam} & \colhead{Channel rms} & \colhead{44 GHz} & \colhead{UC HII\tablenotemark{c}}\\
\colhead{Num \tablenotemark{a}} & \colhead{Num } & \colhead{h m s} & \colhead{\arcdeg\phn\arcmin\phn\arcsec} & \colhead{km s$^{-1}$}   & \colhead{kpc} & \colhead{arcsec} & \colhead{mJy beam$^{-1}$} & \colhead{masers} & \colhead{Outflow}} 
\startdata
  2 H  & 00117+6412   & 00 14 27.7 &+64 28 46   &$-$36.3   & 1.8   &2.37$\times$1.75  &56  & 0 & UO\\
  3 L  & 00420+5530   & 00 44 57.6 &+55 47 18   &$-$51.2   & 7.7   &1.78$\times$1.70  &46  & 0 &  O\\
  7 H  & 04579+4703   & 05 01 39.7 &+47 07 23   &$-$16.5   & 2.4   &1.79$\times$1.63  &53  & 0 &   \\
  8 L  & 05137+3919   & 05 17 13.3 &+39 22 14   &$-$25.4   & 10.8  &1.87$\times$1.51  &48  & 0 &  O\\
  9 H  & 05168+3634   & 05 20 16.2 &+36 37 21   &$-$15.1   & 6.0   &1.89$\times$1.51  &54  & 0 &  O\\
 10 H  & 05274+3345   & 05 30 45.6 &+33 47 52   &$-$3.8    & 1.5   &1.82$\times$1.52  &52  & 8 &  O\\
 11 L  & 05345+3157   & 05 37 47.8 &+31 59 24   &$-$18.4   & 1.8   &1.80$\times$1.48  &58  & 0 &  O\\
 12 L  & 05373+2349   & 05 40 24.4 &+23 50 54   &+2.3      & 1.1   &1.83$\times$1.48  &52  & 0 &  O\\
 15 H  & 06056+2131   & 06 08 41.0 &+21 31 01   &+2.6      & 1.5   &1.88$\times$1.50  &60  & 0 & UO\\
 28 L  & 06584$-$0852 & 07 00 51.5 &$-$08 56 29 &+41.3     & 4.4   &2.72$\times$1.53  &62  & 0 &  O\\
 30 L  & 17450$-$2742 & 17 48 09.3 &$-$27 43 21 &$-$16.6   & 0.1   &3.04$\times$1.33  &82  & 0 &   \\
 36 L  & 18014$-$2428 & 18 04 29.6 &$-$24 28 47 &+13.0     & 2.8   &2.65$\times$1.36  &75  & 0 &   \\
 37 L  & 18018$-$2426 & 18 04 53.9 &$-$24 26 41 &+10.7     & 1.5   &2.61$\times$1.37  &96  & 1 & UO\\
 38 L  & 18024$-$2119 & 18 05 25.4 &$-$21 19 41 &+0.5      & 0.1   &2.33$\times$1.41  &58  & 6 &   \\
 39 L  & 18024$-$2231 & 18 05 30.6 &$-$22 31 36 &+16.1     & 2.9   &2.47$\times$1.38  &67  & 0 &   \\
 45 L  & 18144$-$1723 & 18 17 24.5 &$-$17 22 13 &+47.3     & 4.3   &2.31$\times$1.55  &54  & 11 &  \\
 50 L  & 18162$-$1612 & 18 19 07.5 &$-$16 11 21 &+61.7     & 4.8   &2.28$\times$1.54  &53  & 2 & U\phantom{O}\\
 57 L  & 18256$-$0742 & 18 28 20.5 &$-$07 40 22 &+36.7     & 2.9   &2.03$\times$1.57  &48  & 0 &   \\
 59 L  & 18278$-$1009 & 18 30 35.2 &$-$10 07 12 &+97.7     & 5.7   &2.02$\times$1.60  &44  & 0 &   \\
 60 L  & 18288$-$0158 & 18 31 26.7 &$-$01 56 35 &+3.8      & 0.5   &1.76$\times$1.64  &41  & 0 &   \\
 66 L  & 18363$-$0554 & 18 39 03.7 &$-$05 52 15 &+65.2     & 4.3   &1.88$\times$1.67  &44  & 0 &   \\
 68 L  & 18396$-$0431 & 18 42 18.8 &$-$04 28 37 &+97.4     & 6.0   &1.84$\times$1.66  &48  & 2 & UO\\
 70 L  & 18424$-$0329 & 18 45 03.3 &$-$03 26 49 &+47.9     & 3.3   &1.89$\times$1.69  &44  & 0 &   \\
 74 H  & 18507+0121   & 18 53 17.4 &+01 24 55   &+57.2     & 3.8   &1.83$\times$1.74  &45  & 10 & U\phantom{O}\\
 75 L  & 18511+0146   & 18 53 38.1 &+01 50 27   &+56.5     & 3.8   &1.99$\times$1.40  &39  & 1 &  O\\
 76 H  & 18517+0437   & 18 54 13.8 &+04 41 32   &+43.7     & 3.0   &1.88$\times$1.42  &35  & 3 &   \\
 77 L  & 18527+0301   & 18 55 16.5 &+03 05 07   &+75.8     & 5.2   &1.97$\times$1.36  &46  & 1 &   \\
 78 H  & 18532+0047   & 18 55 50.7 &+00 51 22   &+58.4     & 3.9   &1.91$\times$1.45  &36  & 3 & UO\\
 81 H  & 18551+0302   & 18 57 42.1 &+03 06 04   &+57.0     & 3.9   &1.89$\times$1.46  &41  & 5 & U\phantom{O}\\
 82 L  & 18565+0349   & 18 59 03.4 &+03 53 22   &+91.5     & 6.7   &2.07$\times$1.46  &37  & 1 & U\phantom{O}\\
 84 L  & 18567+0700   & 18 59 13.6 &+07 04 47   &+29.4     & 2.1   &1.77$\times$1.46  &42  & 0 & U\phantom{O}\\
 86 L  & 18571+0349   & 18 59 40.0 &+03 53 35   &+56.2     & 3.8   &1.78$\times$1.47  &38  & 0 &   \\
 87 L  & 18586+0106   & 19 01 10.6 &+01 11 16   &+38.0     & 2.7   &1.89$\times$1.53  &44  & 0 &   \\
 91 L  & 19012+0505   & 19 03 43.5 &+05 09 49   &+40.4     & 2.8   &1.79$\times$1.42  &46  & 0 & U\phantom{O}\\
 93 H  & 19043+0726   & 19 06 47.7 &+07 31 38   &+58.9     & 4.3   &1.71$\times$1.43  &61  & 1 & U\phantom{O}\\
 97 H  & 19088+0902   & 19 11 15.9 &+09 07 27   &+59.6     & 4.7   &1.68$\times$1.39  &59  & 1 & U\phantom{O}\\
 98 L  & 19092+0841   & 19 11 37.4 &+08 46 30   &+58.0     & 4.4   &1.62$\times$1.44  &58  & 5 &   \\
 99 H  & 19094+0944   & 19 11 52.0 &+09 49 46   &+65.3     & 6.1   &1.85$\times$1.45  &65  & 1 &   \\  
102 H  & 19198+1423   & 19 22 07.7 &+14 29 20   &+58.9     & 5.5   &1.66$\times$1.38  &36  & 1 & U\phantom{O}\\
103 H  & 19213+1723   & 19 23 37.0 &+17 28 59   &+41.7     & 4.1   &1.71$\times$1.45  &44  & 0 & UO\\
104 H  & 19282+1814   & 19 30 28.0 &+18 20 53   &+24.1     & 2.1   &1.95$\times$1.53  &171 & 0 &  O\\
108 H  & 19368+2239   & 19 38 58.1 &+22 46 32   &+36.4     & 4.4   &1.92$\times$1.55  &80  & 12 & O\\
109 H  & 19374+2352   & 19 39 33.2 &+23 59 55   &+36.9     & 4.3   &1.91$\times$1.44  &84  & 2 & UO\\
110 H  & 19388+2357   & 19 40 59.4 &+24 04 39   &+34.6     & 4.2   &1.81$\times$1.39  &113 & 2 & UO\\
114 H  & 20050+2720   & 20 07 06.7 &+27 28 53   &+6.4      & 0.7   &1.79$\times$1.49  &89  & 1 & O\\
115 H  & 20056+3350   & 20 07 31.5 &+33 59 39   &+9.4      & 1.6   &1.76$\times$1.45  &83  & 2 & O\\
116 H  & 20062+3550   & 20 08 09.8 &+35 59 20   &+0.6      & 0.0   &1.81$\times$1.39  &113 & 2 & O\\
117 L  & 20099+3640   & 20 11 46.4 &+36 49 37   &$-$36.4   & 8.6   &1.75$\times$1.55  &49  & 0 & UO\\
118 L  & 20106+3545   & 20 12 31.3 &+35 54 46   &+7.8      & 1.6   &1.92$\times$1.64  &51  & 0 & O\\
119 H  & 20126+4104   & 20 14 26.0 &+41 13 32   &$-$3.9    & 4.1   &1.80$\times$1.57  &43  & 5 & O\\
121 H  & 20188+3928   & 20 20 39.3 &+39 37 52   &+1.5      & 0.3   &1.68$\times$1.45  &40  & 1 & UO\\
122 L  & 20217+3947   & 20 23 31.7 &+39 57 23   &$-$0.9    & 3.7   &1.69$\times$1.45  &43  & 0 & \\
123 H  & 20220+3728   & 20 23 55.7 &+37 38 10   &$-$2.7    & 4.4   &1.74$\times$1.52  &46  & 0 & UO\\
125 L  & 20278+3521   & 20 29 46.9 &+35 31 39   &$-$4.5    & 5.0   &1.66$\times$1.40  &42  & 0 & O\\
126 H  & 20286+4105   & 20 30 27.9 &+41 15 48   &$-$3.8    & 3.7   &1.83$\times$1.39  &111 & 2 & O\\
129 L  & 20333+4102   & 20 35 09.5 &+41 13 18   &+8.4      & 0.1   &1.92$\times$1.46  &217 & 0 & U\phantom{O}\\
131 H  & 20444+4629   & 20 46 08.3 &+46 40 41   &$-$4.1    & 2.4   &1.90$\times$1.39  &165 & 0 & U\phantom{O}\\
133 H  & 21078+5211   & 21 09 25.2 &+52 23 44   &$-$6.1    & 1.4   &1.88$\times$1.43  &151 & 0 & UO\\
136 L  & 21307+5049   & 21 32 31.5 &+51 02 22   &$-$46.7   & 4.9   &       -          & -  & \tablenotemark{d} & UO\\
138 H  & 21391+5802   & 21 40 42.4 &+58 16 10   &+0.4      & 0.7   &       -          & -  & \tablenotemark{d} & O\\
139 H  & 21519+5613   & 21 53 39.2 &+56 27 46   &$-$63.2   & 7.3   &       -          & -  & \tablenotemark{d} & O\\
143 L  & 22172+5549   & 22 19 09.0 &+56 04 45   &$-$43.8   & 2.8   &1.77$\times$1.46  &43  & 0 & O\\
145 H  & 22198+6336   & 22 21 27.6 &+63 51 42   &$-$11.3   & 1.2   &1.81$\times$1.43  &49  & 0 & O\\
148 H  & 22305+5803   & 22 32 24.3 &+58 18 58   &$-$51.9   & 5.4   &1.65$\times$1.46  &37  & 0 & O\\
151 H  & 22506+5944   & 22 52 38.6 &+60 00 56   &$-$51.4   & 5.7   &1.65$\times$1.46  &39  & 6 & O\\
154 L  & 23026+5948   & 23 04 45.7 &+60 04 35   &$-$51.1   & 5.7   &2.19$\times$1.42  &84  & 0 & \\
155 L  & 23140+6121   & 23 16 11.7 &+61 37 45   &$-$51.5   & 6.4   &2.19$\times$1.67  &60  & 0 & UO\\
158 L  & 23314+6033   & 23 33 44.4 &+60 50 30   &$-$45.4   & 2.7   &2.30$\times$1.37  &61  & 0 & O\\
160 L  & 23385+6053   & 23 40 53.2 &+61 10 21   &$-$50.0   & 6.9   &2.14$\times$1.43  &59  & 3 & O\\
\enddata
\tablenotetext{a}{The letters H and L denote the source classification of {\it High} or {\it Low} by M96.}
\tablenotetext{b}{Kinematic distances from Molinari et al. (1998).}
\tablenotetext{c}{The letter U indicates that centimeter radio continuum was detected in the VLA maps of Molinari et al. (1998), Palau et al. (2010), or Kurtz, Churchwell, \& Wood (1994), indicating a UC HII region candidate.  The letter O indicates that Zhang et al. (2005) reported a molecular outflow in this source. If blank, then neither continuum emission nor an outflow was detected.}
\tablenotetext{d}{For these sources the phase calibrator was too weak to provide usable solutions.}

\end{deluxetable}

\clearpage
\begin{deluxetable}{lcrrrrcrc}
\tablecolumns{9}
\tabletypesize{\footnotesize}
\tablewidth{0pt}
\tablenum{2}
\tablecaption{44~GHz Masers\label{tbl-2}}
\tablehead{
\colhead{} & \colhead{} & \multicolumn{2}{c}{Maser Peak Position} & \colhead{} & \colhead{} 
& \colhead{} & \colhead{} & \colhead{}\\
\colhead{Source} & \colhead{Maser} & \colhead{$\alpha$(J2000)} & \colhead{$\delta$(J2000)} & \colhead{$v_{LSR}$} & \colhead{$S_{\nu}$} 
& \colhead{FWZI} & \colhead{$\int S_{\nu}dv$} & \colhead{}\\
\colhead{IRAS (Mol)} & \colhead {Number} & \colhead{h m s} & \colhead{\arcdeg\phn\arcmin\phn\arcsec} & \colhead{km s$^{-1}$} & \colhead{Jy} 
& \colhead{km s$^{-1}$} & \colhead{Jy km s$^{-1}$} & \colhead{Note} } 

\startdata
05274+3345 (10) &  1 &  05 30 47.17 & $+$33 47 58.1&   $-$3.0        &            0.61&    0.83 &        0.28&\\
           &  2 &  05 30 47.53 & $+$33 47 56.1&   $-$2.8        &            2.00&    1.83 &        1.18&\\
           &  3 &  05 30 47.62 & $+$33 47 52.2&   $-$3.5        &            2.93&    1.16 &        0.84&   \\
           &  4 &  05 30 47.62 & $+$33 47 52.2&   $-$4.3        &            0.44&    - &        -&   \tablenotemark{a}\\
           &  5 &  05 30 47.68 & $+$33 47 56.5&   $-$3.1        &            0.53&    1.49 &        0.27&   \\
           &  6 &  05 30 47.68 & $+$33 47 56.5&   $-$3.7        &            0.15&    - &        -&   \tablenotemark{a}\\
           &  7 &  05 30 47.80 & $+$33 47 55.3&   $-$2.8        &            0.49&    0.66 &        0.15&\\
           &  8 &  05 30 47.98 & $+$33 47 56.3&   $-$3.0        &            0.40&    0.83 &        0.22&\\
18018-2426 (37) &  1 &  18 04 53.01 & $-$24 26 40.5&  $+$10.9        &          579.41&    1.16 &      329.51&\\
18024-2119 (38) &  1 &  18 05 24.94 & $-$21 19 15.4&   $-$1.0        &            0.69&    1.00 &        0.42&\\
           &  2 &  18 05 25.39 & $-$21 19 16.9&       0.0         &           11.19&    1.49 &        6.44&\\
           &  3 &  18 05 25.86 & $-$21 19 24.8&   $-$0.3        &            2.67&    1.83 &        0.57&\\
           &  4 &  18 05 25.86 & $-$21 19 24.8&   $+$1.8        &            0.36&    0.17 &        0.01&\\
           &  5 &  18 05 25.89 & $-$21 19 23.6&   $-$0.8        &            2.79&    0.66 &        1.01&    \\
           &  6 &  18 05 26.10 & $-$21 19 27.2&   $+$2.5        &            0.56&    0.50 &        0.19&\\
18144-1723 (45) &  1 &  18 17 22.93 & $-$17 22 13.5&  $+$48.3        &            6.82&    2.99 &        5.55&\\
           &  2 &  18 17 23.09 & $-$17 22 14.4&  $+$49.0        &            5.27&    1.33 &        1.05&    \\
           &  3 &  18 17 23.09 & $-$17 22 14.4&  $+$49.7        &            0.14&    - &        -&    \tablenotemark{a}\\
           &  4 &  18 17 23.09 & $-$17 22 17.8&  $+$48.6        &            0.28&    1.99 &        0.33& \\
           &  5 &  18 17 23.10 & $-$17 22 17.4&  $+$47.5        &            0.59&      -  &          - & \tablenotemark{a}\\
           &  6 &  18 17 23.22 & $-$17 22 14.7&  $+$47.6        &            1.05&    3.15 &        1.48& \\
           &  7 &  18 17 23.43 & $-$17 22 10.9&  $+$46.6        &            0.39&      -  &          - & \tablenotemark{a}\\
           &  8 &  18 17 23.34 & $-$17 22 11.3&  $+$47.3        &            9.37&    1.99 &        3.64& \\
           &  9 &  18 17 23.34 & $-$17 22 11.3&  $+$46.0        &            1.28&      -  &          - & \tablenotemark{a}\\
           &  10 &  18 17 24.07 & $-$17 22 13.8&  $+$48.5        &           45.14&    2.66 &       16.31& \\
           & 11 &  18 17 24.06 & $-$17 22 13.8&  $+$48.0        &           14.34&      -  &          - & \tablenotemark{a}\\
18162-1612 (50) &  1 &  18 19 07.60 & $-$16 11 24.6&  $+$64.0        &            1.09&    0.66 &       0.46 &\\
           &  2 &  18 19 07.63 & $-$16 11 25.4&  $+$62.2        &           11.19&    1.16 &       6.82 &\\
18396-0431 (68) &  1 &  18 42 17.91 & $-$04 28 56.5&  $+$97.4        &            0.34&    0.83 &       0.52 &\\
           &  2 &  18 42 17.88 & $-$04 28 55.6&       $+$97.1   &            0.34&       - &          - & \tablenotemark{a}\\
18507+0121 (74) &  1 &  18 53 17.30 & $+$01 24 42.4 & $+$54.9        &           1.40 &    1.66 &       0.98 &\\
           &  2 &  18 53 17.30 & $+$01 24 42.4 & $+$55.5        &           0.82 &       - &          - & \tablenotemark{a}\\
           &  3 &  18 53 18.46 & $+$01 24 50.6 & $+$58.2        &           1.33 &    1.16 &       0.51 &\\
           &  4 &  18 53 18.68 & $+$01 24 41.2 & $+$60.2        &           21.12&    2.66 &       7.76 &\\
           &  5 &  18 53 18.69 & $+$01 24 41.2 & $+$59.7        &           6.85 &      -  &         -  & \tablenotemark{a} \\
           &  6 &  18 53 18.73 & $+$01 24 31.1 & $+$58.9        &           5.69 &    3.15 &       4.55 &\\
           &  7 &  18 53 19.07 & $+$01 24 22.6 & $+$57.5        &           0.69 &    0.33 &       0.15 &\\
           &  8 &  18 53 19.05 & $+$01 24 27.9 & $+$57.7        &           2.65 &    0.50 &       0.75 &    \\
           &  9 &  18 53 19.06 & $+$01 24 33.4 & $+$59.7        &           3.42 &    1.33 &       2.05 &\\
           & 10 &  18 53 19.13 & $+$01 24 21.7 & $+$58.7        &           0.42 &    0.33 &       0.08 &\\
18511+0146 (75) &  1 &  18 53 37.71 & $+$01 50 25.5&  $+$56.3        &           1.14 &    0.50 &       0.39 &\\
18517+0437 (76) &  1 &  18 54 13.81 & $+$04 41 34.0&  $+$42.2        &           0.33 &    0.33 &       0.09 &\\
           &  2 &  18 54 14.46 & $+$04 41 44.5&  $+$44.2        &           1.28 &    1.83 &       1.05 &    \\
           &  3 &  18 54 14.73 & $+$04 41 42.9&  $+$43.5        &           3.32 &    1.00 &       2.30 &\\
18527+0301 (77) &  1 &  18 55 16.77 & $+$03 05 06.9&  $+$75.5        &           1.00 &    0.33 &       0.27 &\\
18532+0047 (78) &  1 &  18 55 51.18 & $+$00 51 12.3&  $+$56.9        &           0.66 &    0.33 &       0.20 &\\
           &  2 &  18 55 51.34 & $+$00 51 26.4&  $+$60.6        &           0.81 &    1.33 &       0.66 &     \\
           &  3 &  18 55 51.35 & $+$00 51 26.4&  $+$62.2        &           0.37 &    0.83 &       0.24 &     \\
18551+0302 (81) &  1 &  18 57 41.57 & $+$03 06 02.9&  $+$56.0        &           0.71 &    0.50 &       0.24 &\\
           &  2 &  18 57 41.81 & $+$03 06 02.7&  $+$56.0        &           2.88 &    0.83 &       1.17 &     \\
           &  3 &  18 57 41.81 & $+$03 06 02.7&  $+$54.7        &           0.91 &    1.00 &       0.58 &     \\
           &  4 &  18 57 41.87 & $+$03 06 09.2&  $+$56.3        &           0.54 &    0.50 &       0.19 &     \\
           &  5 &  18 57 41.87 & $+$03 06 09.6&  $+$55.3        &           0.41 &    0.50 &       0.16 &     \\
18565+0349 (82) &  1 &  18 59 03.73 & $+$03 53 42.8&  $+$90.2        &           1.37 &    1.00 &       0.74 &\\
19043+0726 (93) &  1 &  19 06 47.84 & $+$07 31 41.9&  $+$58.2        &           8.92 &    1.66 &       0.12 &\\
19088+0902 (97) &  1 &  19 11 17.34 & $+$09 07 33.1&  $+$58.1        &          12.91 &    1.83 &       5.13 &   \\
19092+0841 (98) &  1 &  19 11 38.76 & $+$08 46 37.9&  $+$58.0        &           5.68 &    0.50 &       1.71 &     \\
           &  2 &  19 11 38.78 & $+$08 46 38.1&  $+$57.5        &           0.49 &    0.66 &       0.32 &     \\
           &  3 &  19 11 38.77 & $+$08 46 33.1&  $+$57.3        &           1.24 &    1.00 &       0.65 &\\
           &  4 &  19 11 39.08 & $+$08 46 34.2&  $+$57.3        &           0.30 &    0.33 &       0.07 &\\
           &  5 &  19 11 39.12 & $+$08 46 30.7&  $+$56.2        &           0.53 &    0.66 &       0.23 &\\
19094+0944 (99) &  1 &  19 11 51.47 & $+$09 49 41.7&  $+$66.0        &           0.62 &    0.83 &       0.28 &\\
19198+1423 (102) &  1 &  19 22 08.10 & $+$14 29 17.2&  $+$57.2        &           0.35 &    0.33 &       0.09 &\\
19368+2239 (108) &  1 &  19 38 56.31 & $+$22 46 29.3 & $+$38.6        &	    1.56 &    2.0  & 	   3.03 &\\	
           &  2	&  19 38 56.30 & $+$22 46 29.4 & $+$36.7        &	    1.23 &    3.0  &       3.03 &\\
	   &  3	&  19 38 56.74 & $+$22 46 31.3 & $+$36.1        &           5.64 &    1.16 &	   1.92 &\\	
	   &  4	&  19 38 56.82 & $+$22 46 33.1 & $+$37.9        &          18.55 &    1.00 &	   7.56 &\\ 
	   &  5	&  19 38 57.24 & $+$22 46 34.8 & $+$36.6        &           0.73 &    1.49 &	   0.70 &\\	
	   &  6	&  19 38 57.34 & $+$22 46 37.3 & $+$35.1        &           0.71 &    0.33 &	   0.13 &\\	
	   &  7	&  19 38 57.45 & $+$22 46 37.7 & $+$36.9        &           0.56 &    0.66 &	   0.20 &\\
	   &  8	&  19 38 57.47 & $+$22 46 35.8 & $+$35.6        &           0.58 &    0.16 &	   0.07 &\\	
	   &  9 &  19 38 57.74 & $+$22 46 35.7 & $+$35.9        &           1.33 &    0.66 &	   0.45 &\\
	   &  10&  19 38 57.85 & $+$22 46 36.9 & $+$37.2        &           0.48 &    0.50 &	   0.15 &\\
	   &  11&  19 38 58.14 & $+$22 46 37.3 & $+$38.2        &           0.67 &    0.50 &	   0.23 &\\
	   &  12&  19 38 59.18 & $+$22 46 46.0 & $+$36.1        &           1.30 &    0.16 &	   0.19 &\\
19374+2352 (109) &  1	&  19 39 35.10 & $+$23 59 43.1 & $+$34.2        &	    1.32 &    0.66 &	   0.58 &\\
	   &  2	&  19 39 35.17 & $+$23 59 39.1 & $+$34.6        &	    0.53 &    0.33 &	   0.13 &\\
19388+2357 (110) &  1 &  19 40 59.04 & $+$24 04 44.2 & $+$35.1        &	    1.77 &    1.33 &	   1.53 & \\
	   &  2 &  19 40 59.06 & $+$24 04 44.5 & $+$34.4        &	    0.61 &      -  &         -  & \tablenotemark{a}\\
20050+2720 (114) &  1	&  20 07 06.20 & $+$27 28 58.5 & $+$7.4         &	    1.16 &    0.16 &	   0.17 &\\
20056+3350 (115) &  1	&  20 07 31.51 & $+$33 59 46.9 & $+$8.9         &	    0.76 &    0.33 &	   0.19 &\\
	   &  2	&  20 07 31.60 & $+$33 59 44.7 & $+$10.1        &	    0.51 &    0.33 &	   0.14 &\\
20062+3550 (116) &  1	&  20 08 10.23 & $+$35 59 29.0 & $+$0.2        &	   12.04 &    1.33 &	   7.42 &\\
	   &  2	&  20 08 10.33 & $+$35 59 23.5 & $-$0.3        &	    1.29 &    0.50 &	   0.35 &\\
20126+4104 (119) &  1 &  20 14 25.18 & $+$41 13 36.1&  $-$2.4         &           2.47 &    2.16 &       1.75 &\\
           &  2 &  20 14 25.22 & $+$41 13 34.9&  $-$2.4         &           2.16 &    1.33 &       1.14 &\\
           &  3 &  20 14 25.31 & $+$41 13 40.6&  $-$3.1         &           1.14 &    0.33 &       0.30 &\\
           &  4 &  20 14 25.42 & $+$41 13 37.7&  $-$3.2         &           4.24 &    1.49 &       2.36 &\\
           &  5 &  20 14 26.72 & $+$41 13 29.7&  $-$4.4         &           2.84 &    0.83 &       1.20 &\\
20188+3928 (121) &  1 &  20 20 39.79 & $+$39 38 15.3&  $+$2.3         &           1.88 &    1.33 &       1.23 &\\
20286+4105 (126) &  1	&  20 30 28.98 & $+$41 15 47.2 & $-$3.5         &	    1.07 &    0.16 &	   0.13 &\\ 
	   &  2	&  20 30 29.16 & $+$41 15 49.5 & $-$4.0         &	    6.29 &    1.00 &	   2.67 &\\
22506+5944 (151) &  1 &  22 52 37.94 & $+$60 00 46.6&  $-$51.2        &           0.52 &    0.33 &       0.12 &\\
           &  2 &  22 52 38.13 & $+$60 00 40.0&  $-$50.7        &           0.59 &    1.16 &       0.38 &    \\
           &  3 &  22 52 38.34 & $+$60 00 47.2&  $-$51.1        &           1.37 &    0.83 &       0.50 &\\
           &  4 &  22 52 38.83 & $+$60 00 44.1&  $-$52.6        &           1.03 &    0.33 &       0.33 &     \\
           &  5 &  22 52 39.66 & $+$60 00 44.0&  $-$52.2        &           0.54 &    0.50 &       0.09 &\\
           &  6 &  22 52 39.86 & $+$60 00 41.7&  $-$51.7        &           2.54 &    1.49 &       1.69 &\\
23385+6053 (160)  &  1 &  23 40 54.46 & $+$61 10 30.8 &  $-$49.7 &   0.47 &   0.34 &   0.21 &\\
                  &  2 &  23 40 54.55 & $+$61 10 30.9 &  $-$50.5 &   0.52 &   0.85 &   0.25 &\\
                  &  3 &  23 40 54.59 & $+$61 10 29.0 &  $-$52.2 &   0.63 &   0.68 &   0.60 & \\  
                  &  4 &  23 40 54.81 & $+$61 10 28.8 &  $-$51.7 &   0.26 &   -    &    -  & \tablenotemark{a}\\

\enddata
\tablenotetext{a}{Overlapping feature: the velocity and intensity of the secondary peak is reported, but linewidth and integrated flux is reported as part of the main component in the line above.}
\end{deluxetable}

\clearpage

\begin{figure}
\epsscale{0.85}
  \begin{tabular}{@{}cc@{}}
\includegraphics[width=.47\textwidth,angle=-90]{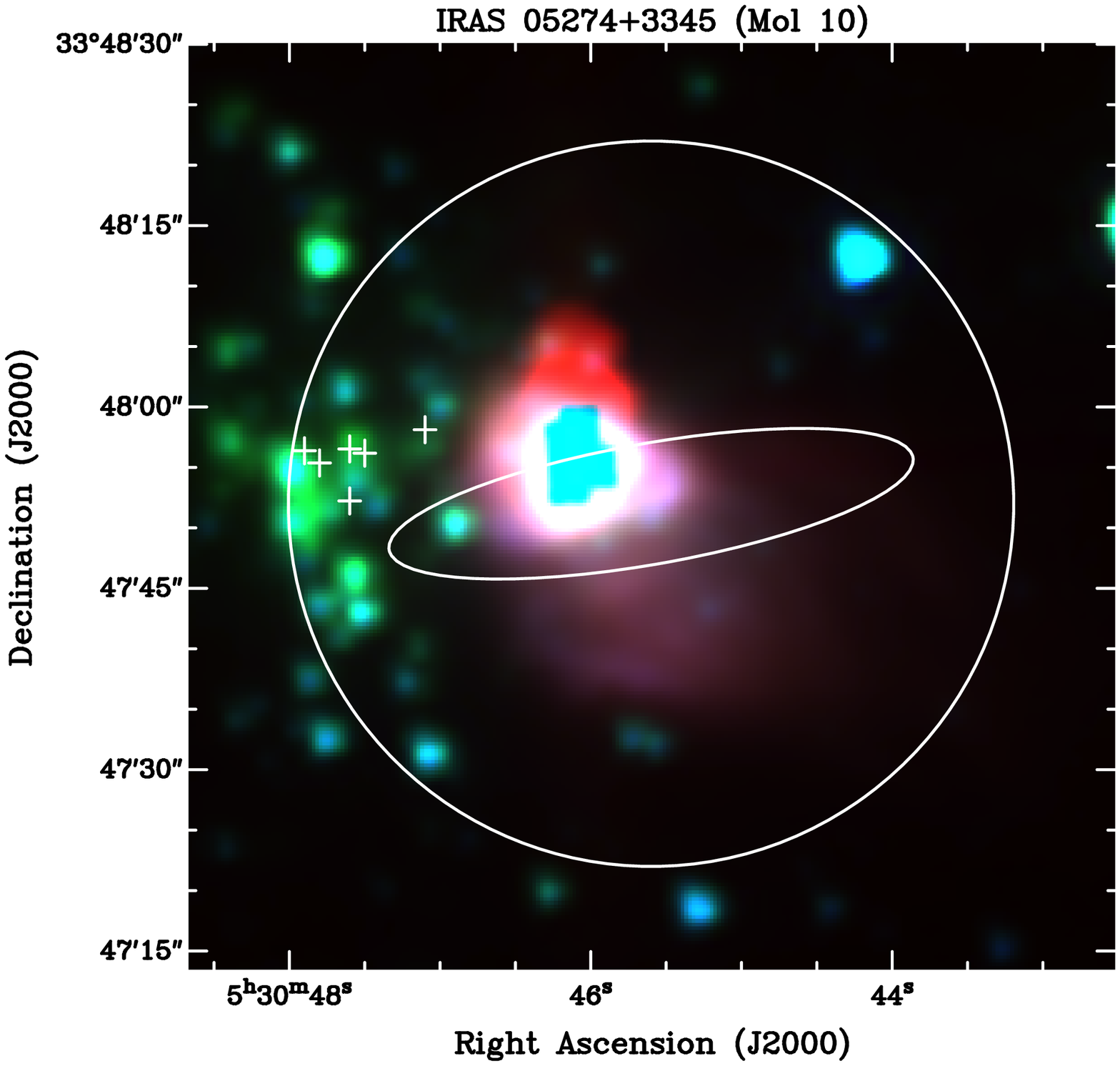} & 
\includegraphics[width=.47\textwidth,angle=-90]{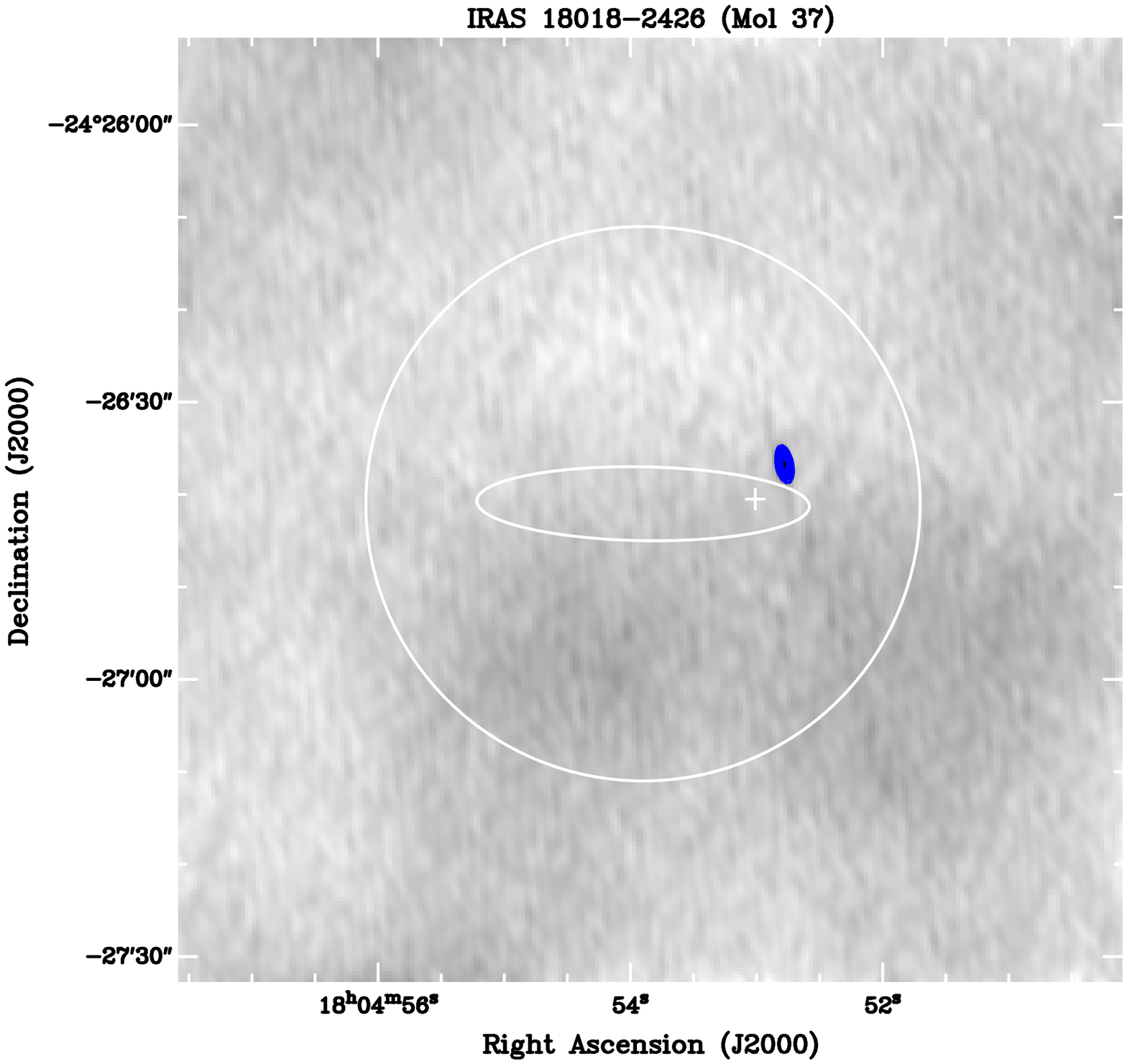} \\
\includegraphics[width=.47\textwidth,angle=-90]{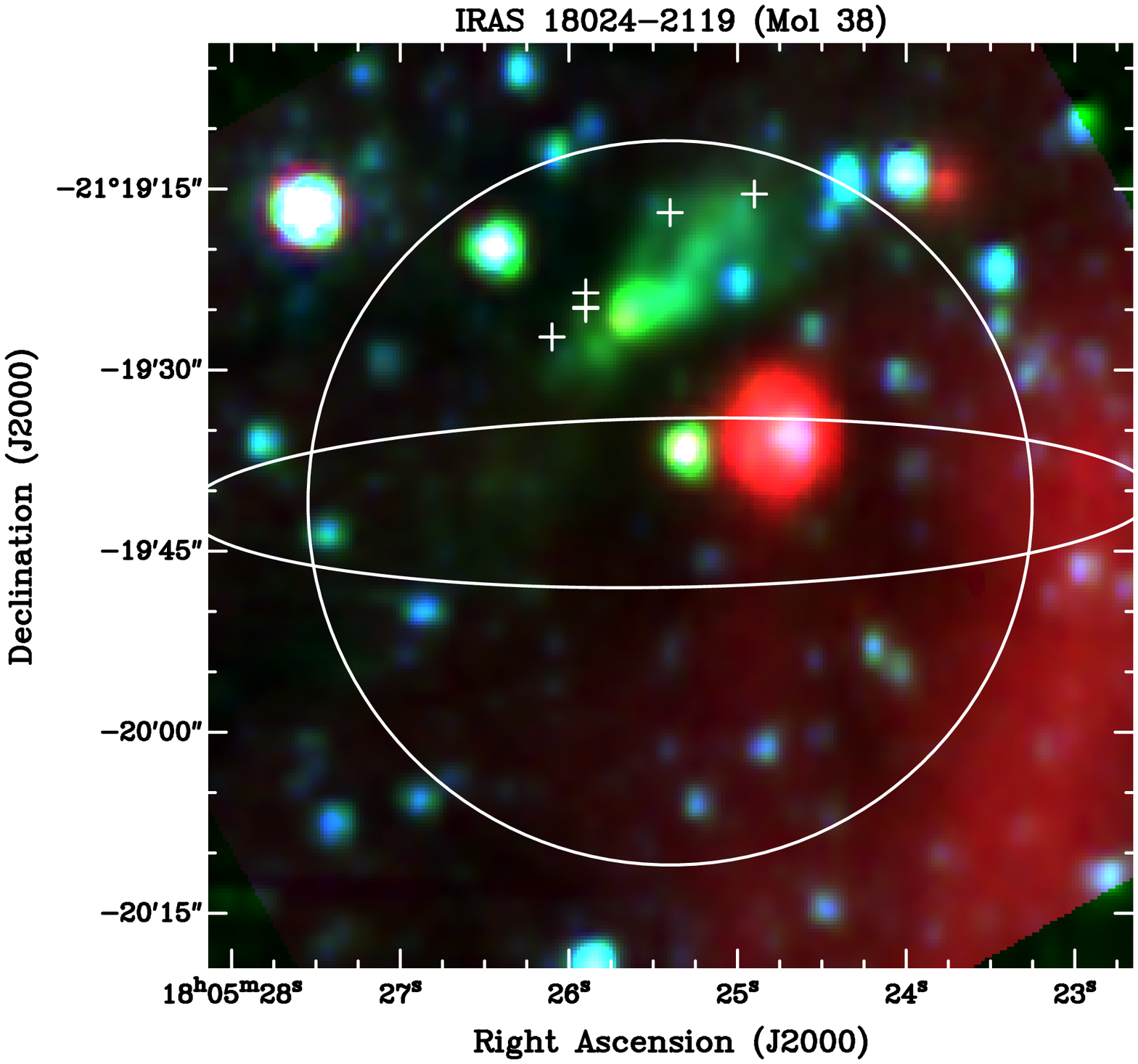} &
\includegraphics[width=.47\textwidth,angle=-90]{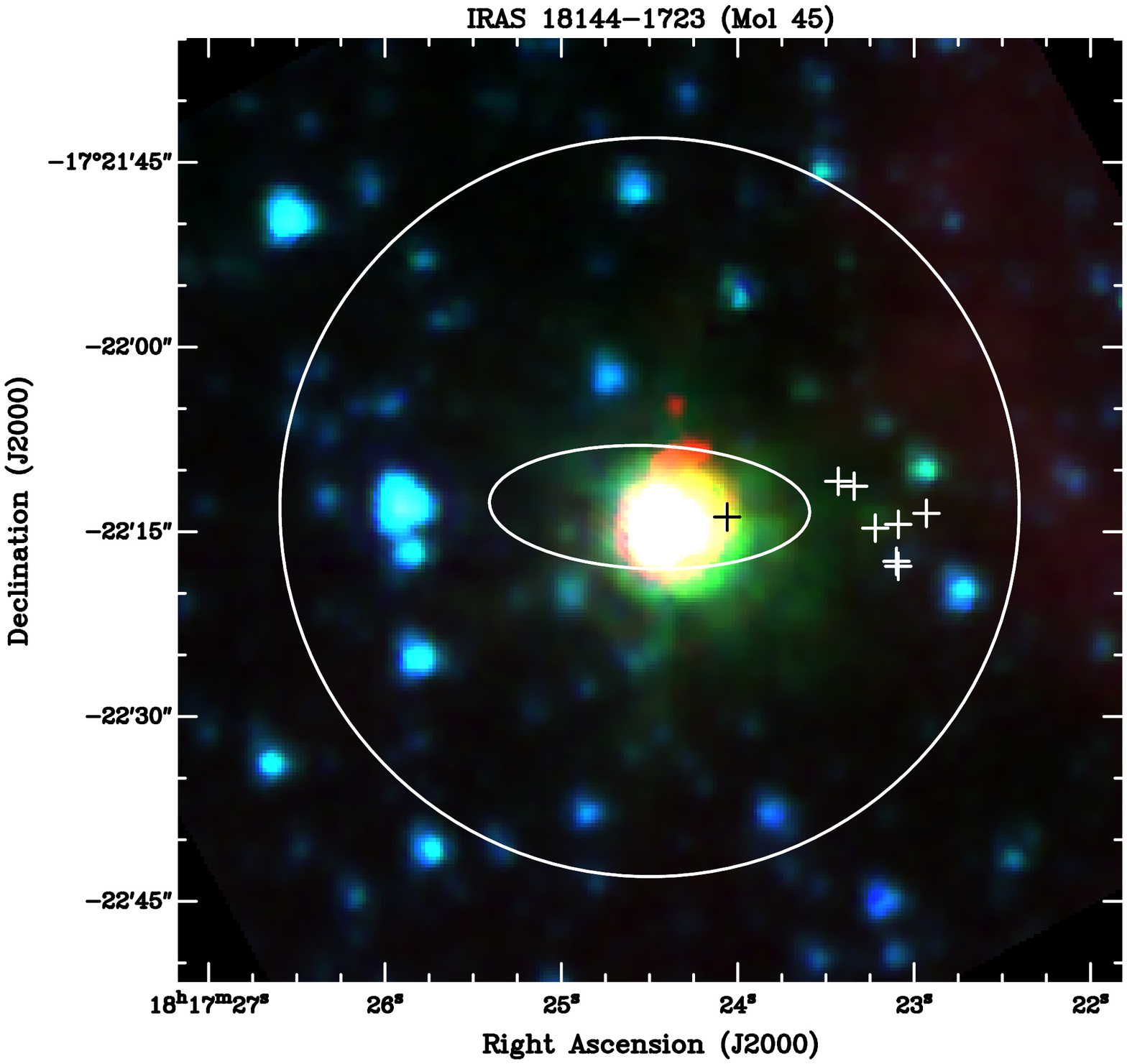} \\
\end{tabular}
\caption{Three-color images from IRAC/Spitzer (3.6$\mu$m: blue, 4.5$\mu$m: green, 8.0$\mu$m: red). 44 GHz methanol masers are represented by plus symbols, the IRAS position is indicated by the ellipse, while the VLA primary beam by a circle. When available, the 3.6 cm continuum from Molinari et al. (1998) is shown in blue contours (in steps of 10\% of the peak). When only the 3.6 cm continuum is available, the emission is shown in grey scale and blue contours. 
\label{3col}}
\end{figure}

\addtocounter{figure}{-1}
\begin{figure}
\epsscale{0.85}
  \begin{tabular}{@{}cc@{}}
\includegraphics[width=.47\textwidth,angle=-90]{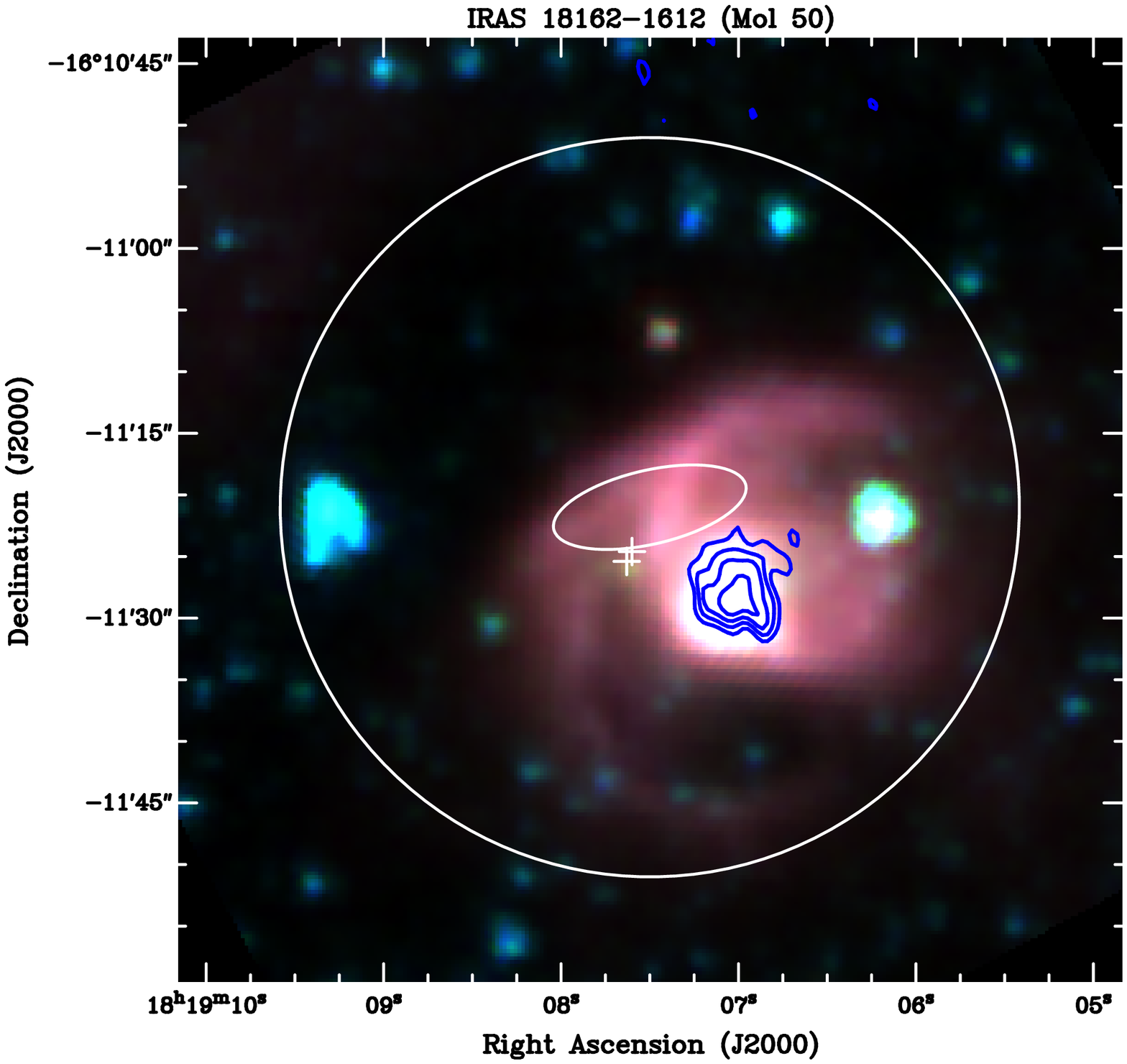} & 
\includegraphics[width=.47\textwidth,angle=-90]{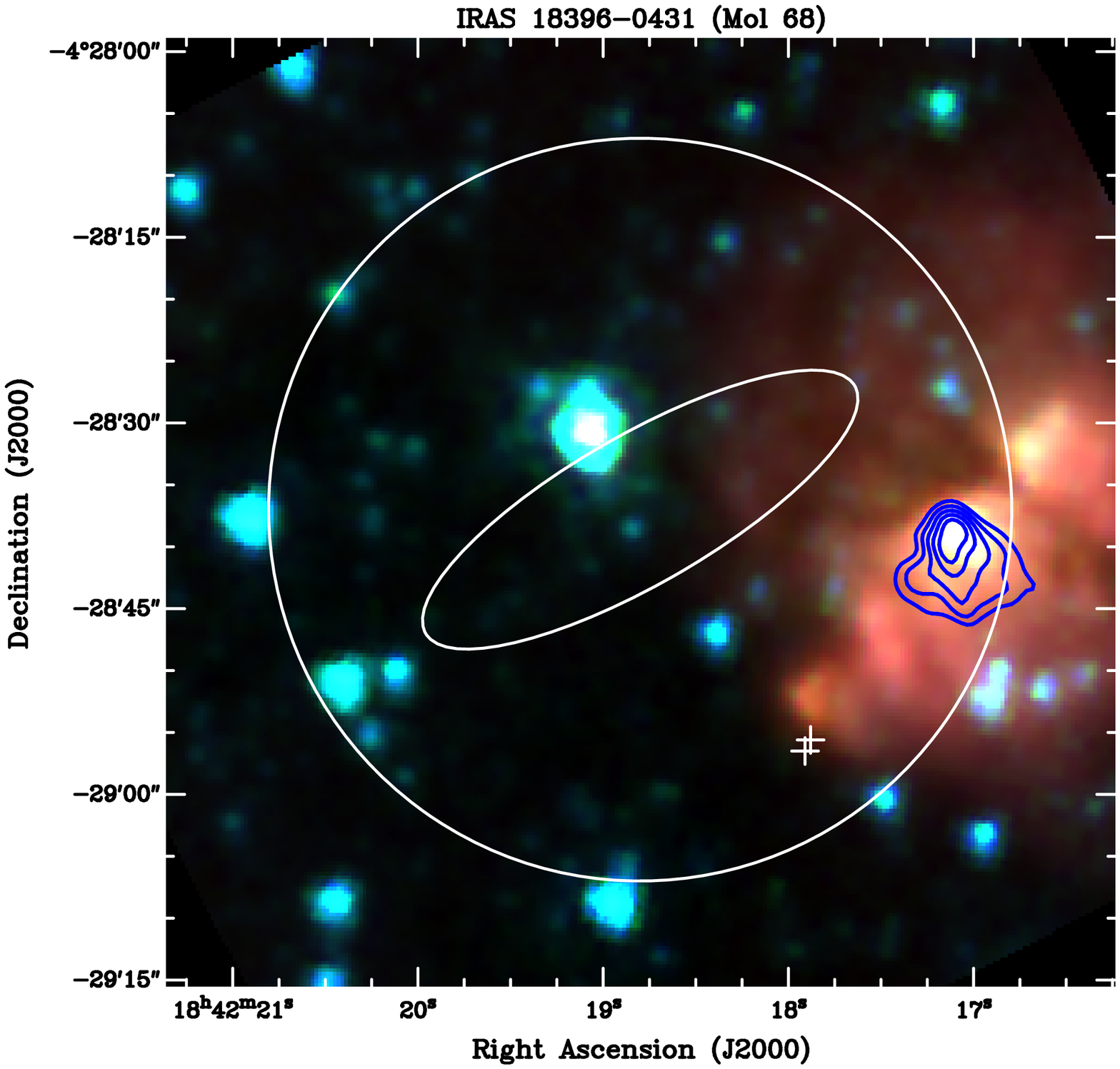} \\
\includegraphics[width=.47\textwidth,angle=-90]{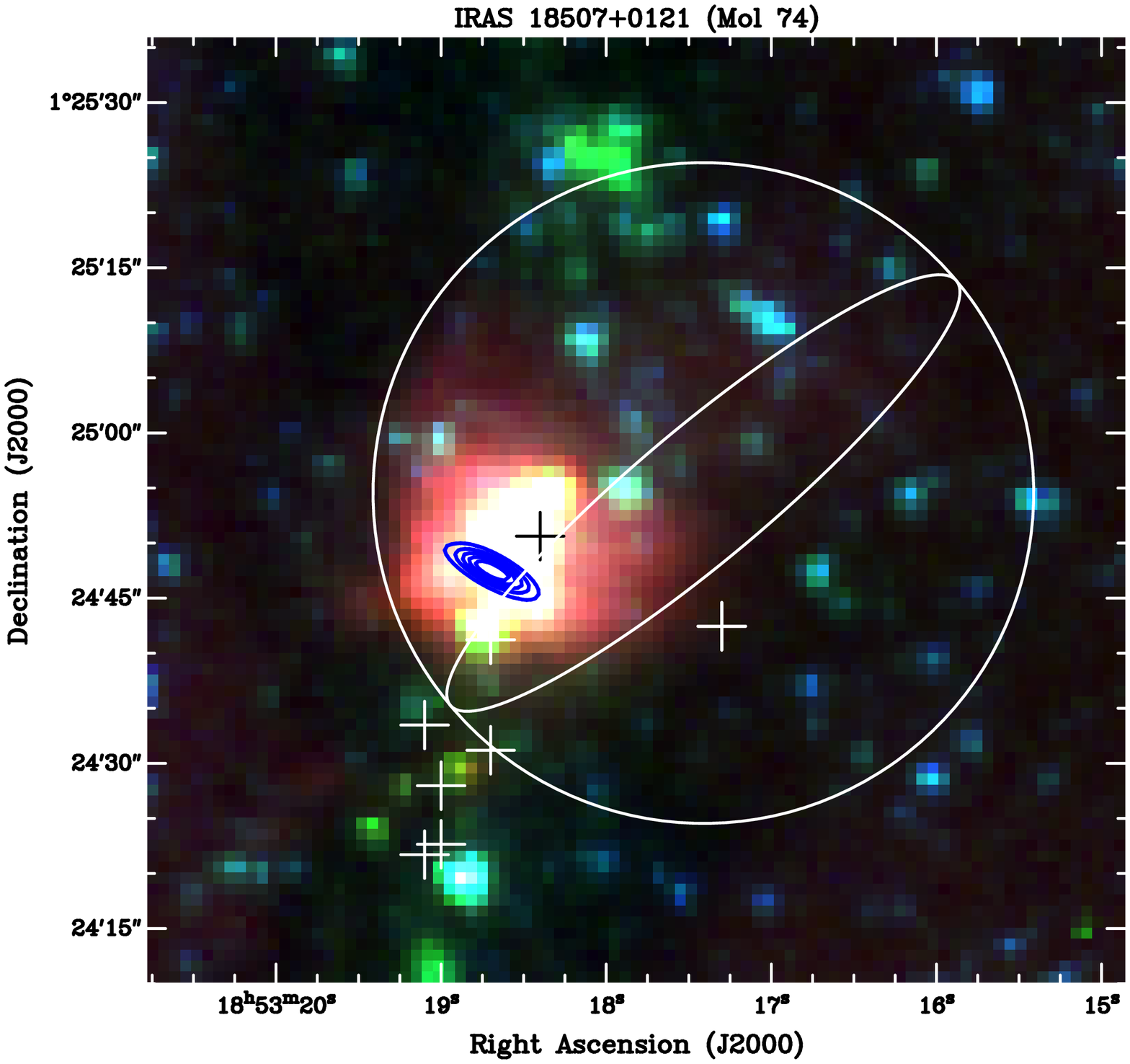} &
\includegraphics[width=.47\textwidth,angle=-90]{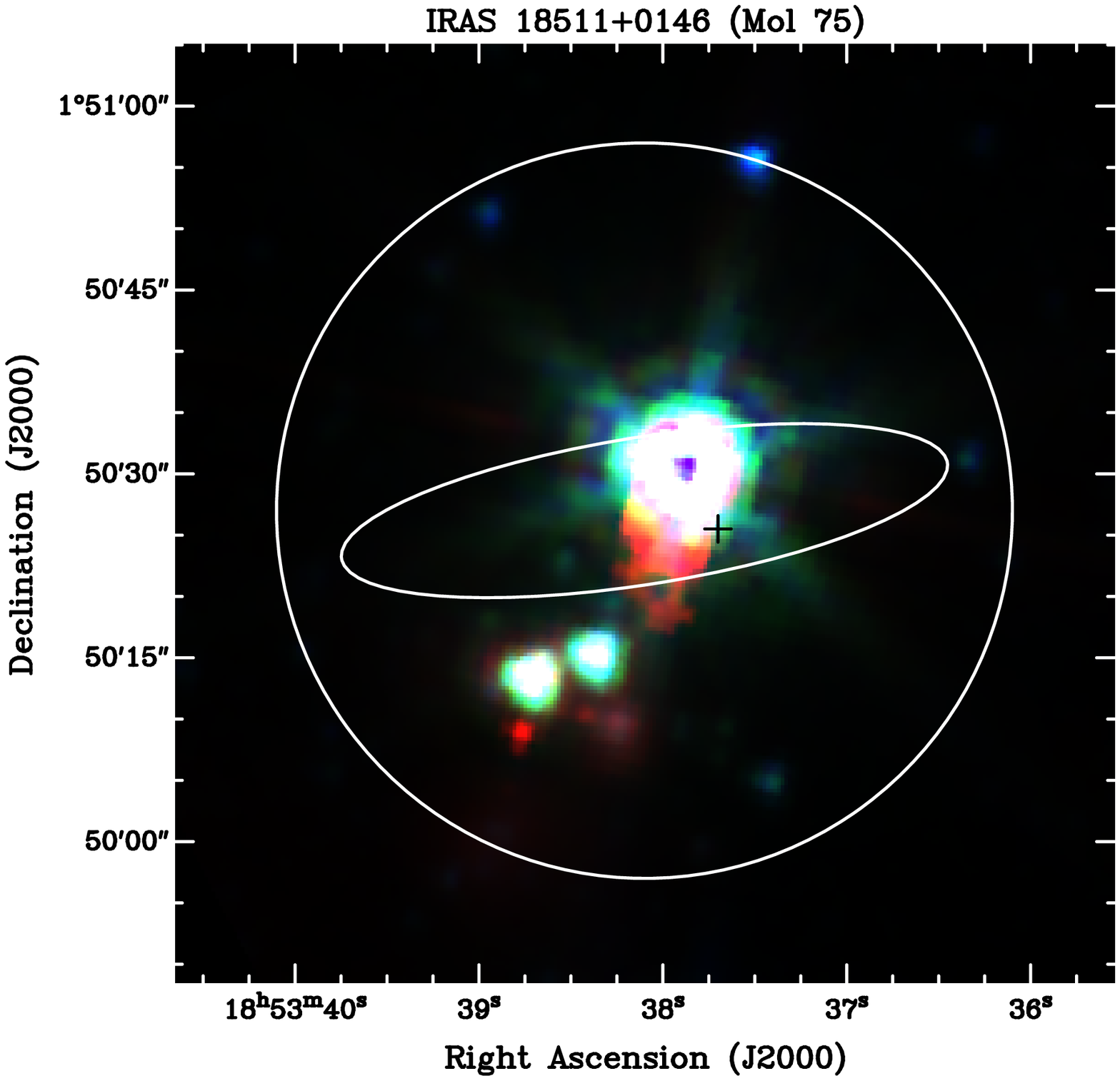} \\
\end{tabular}
\caption{Cont.
\label{3col-2}}
\end{figure}

\addtocounter{figure}{-1}
\begin{figure}
\epsscale{0.85}
  \begin{tabular}{@{}cc@{}}
\includegraphics[width=.47\textwidth,angle=-90]{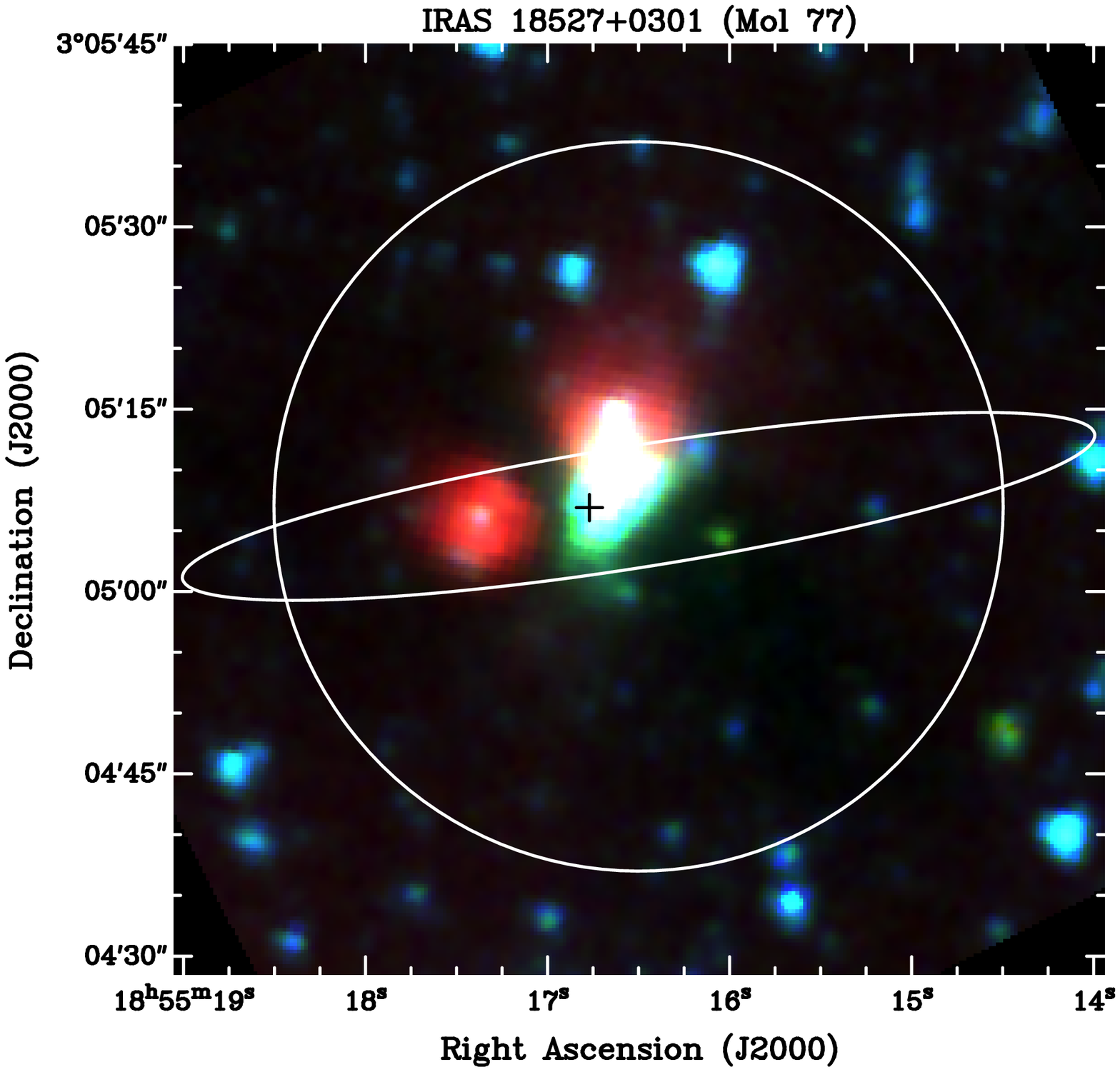} & 
\includegraphics[width=.47\textwidth,angle=-90]{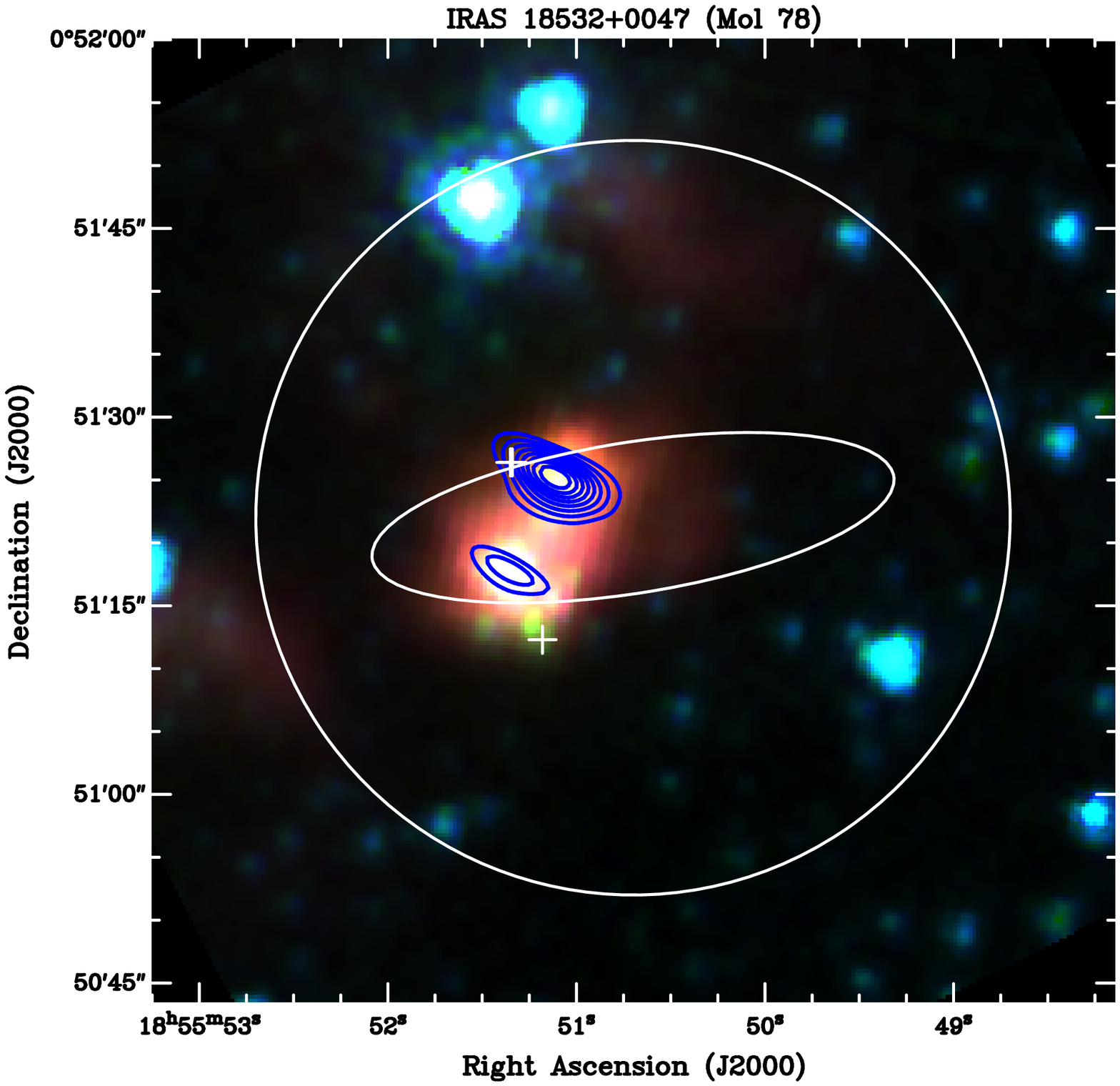} \\
\includegraphics[width=.47\textwidth,angle=-90]{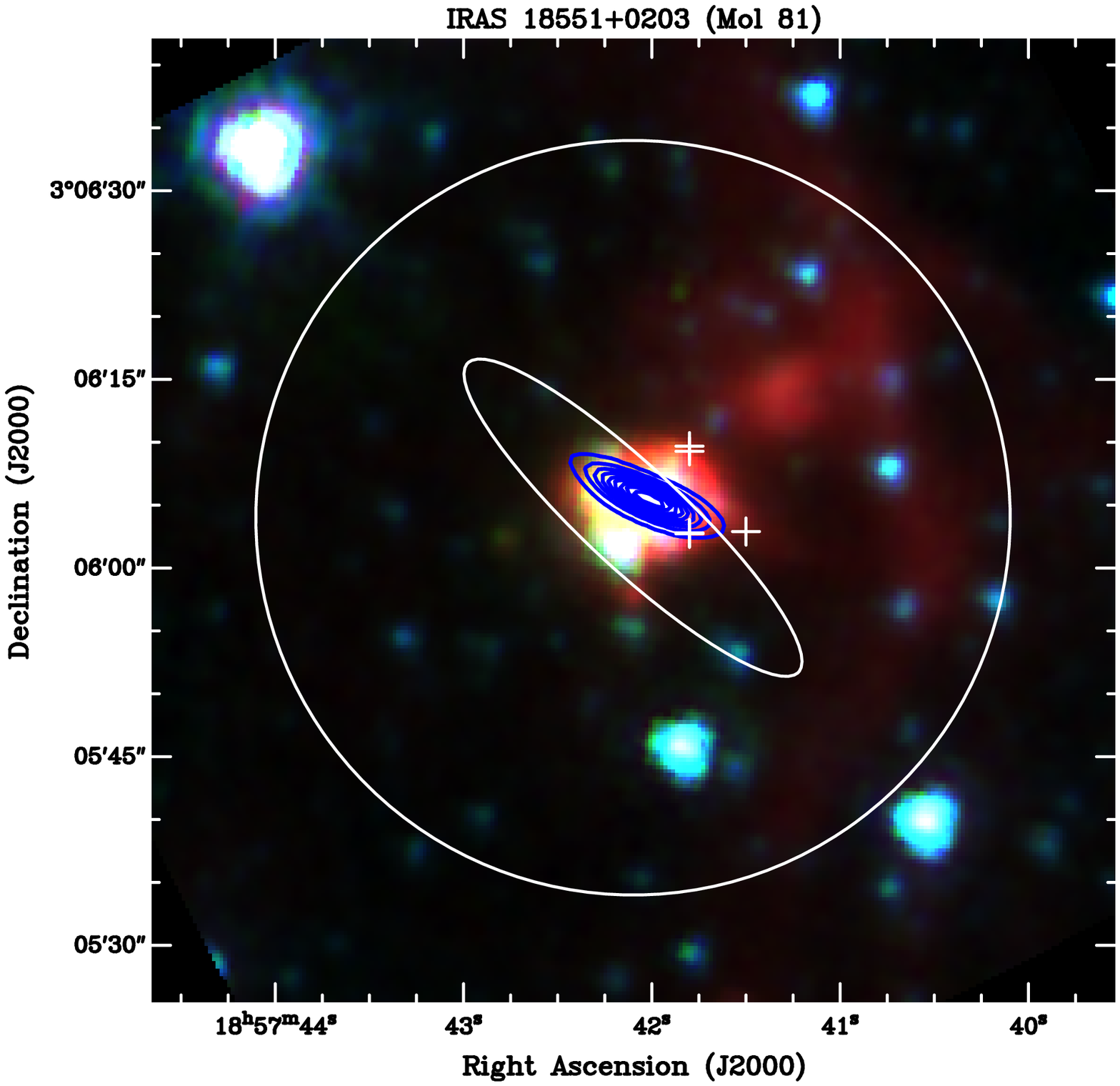} &
\includegraphics[width=.47\textwidth,angle=-90]{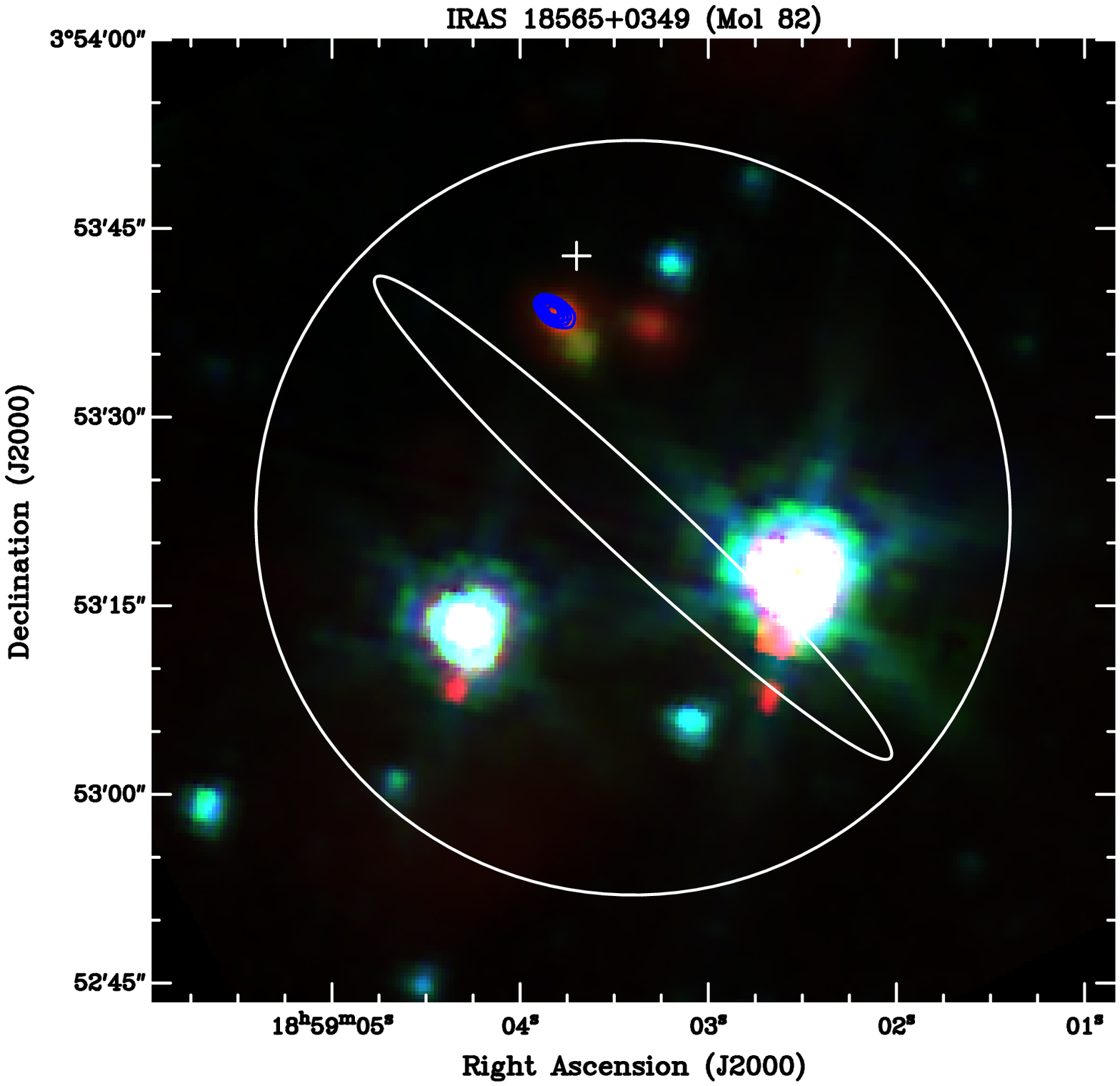} \\
\end{tabular}
\caption{Cont.
\label{3col-3}}
\end{figure}

\addtocounter{figure}{-1}
\begin{figure}
\epsscale{0.85}
  \begin{tabular}{@{}cc@{}}
\includegraphics[width=.47\textwidth,angle=-90]{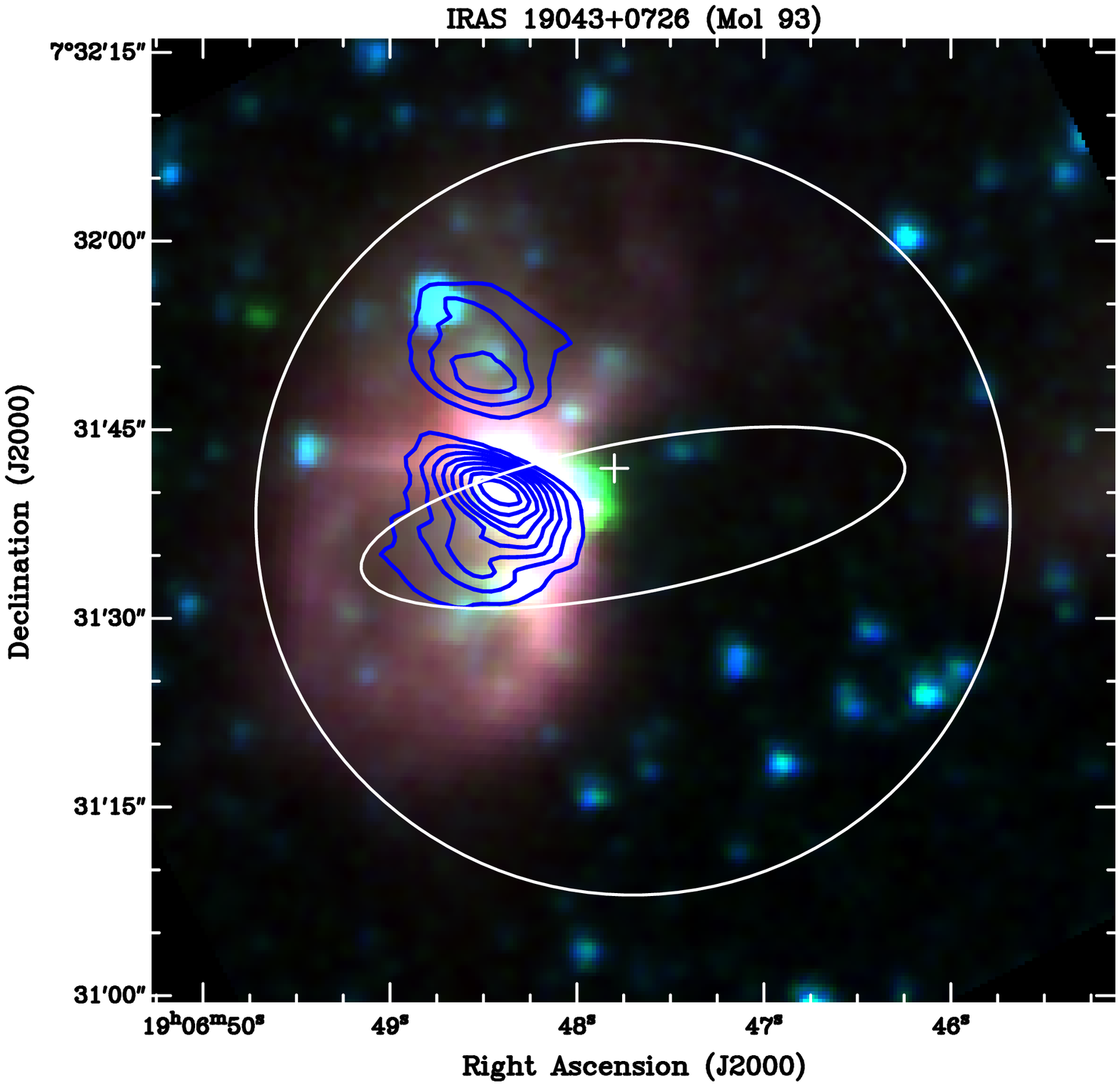} & 
\includegraphics[width=.47\textwidth,angle=-90]{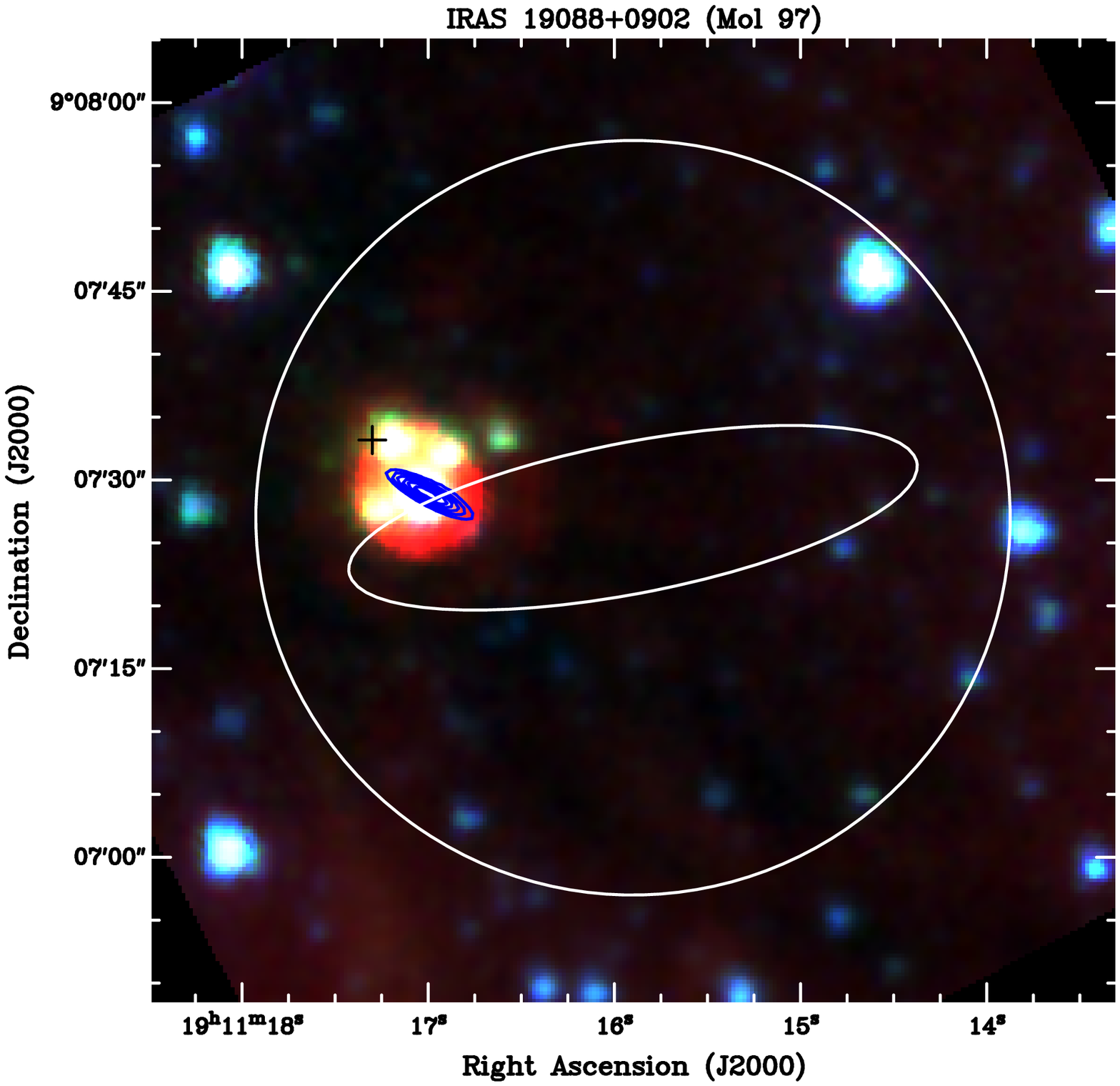} \\
\includegraphics[width=.47\textwidth,angle=-90]{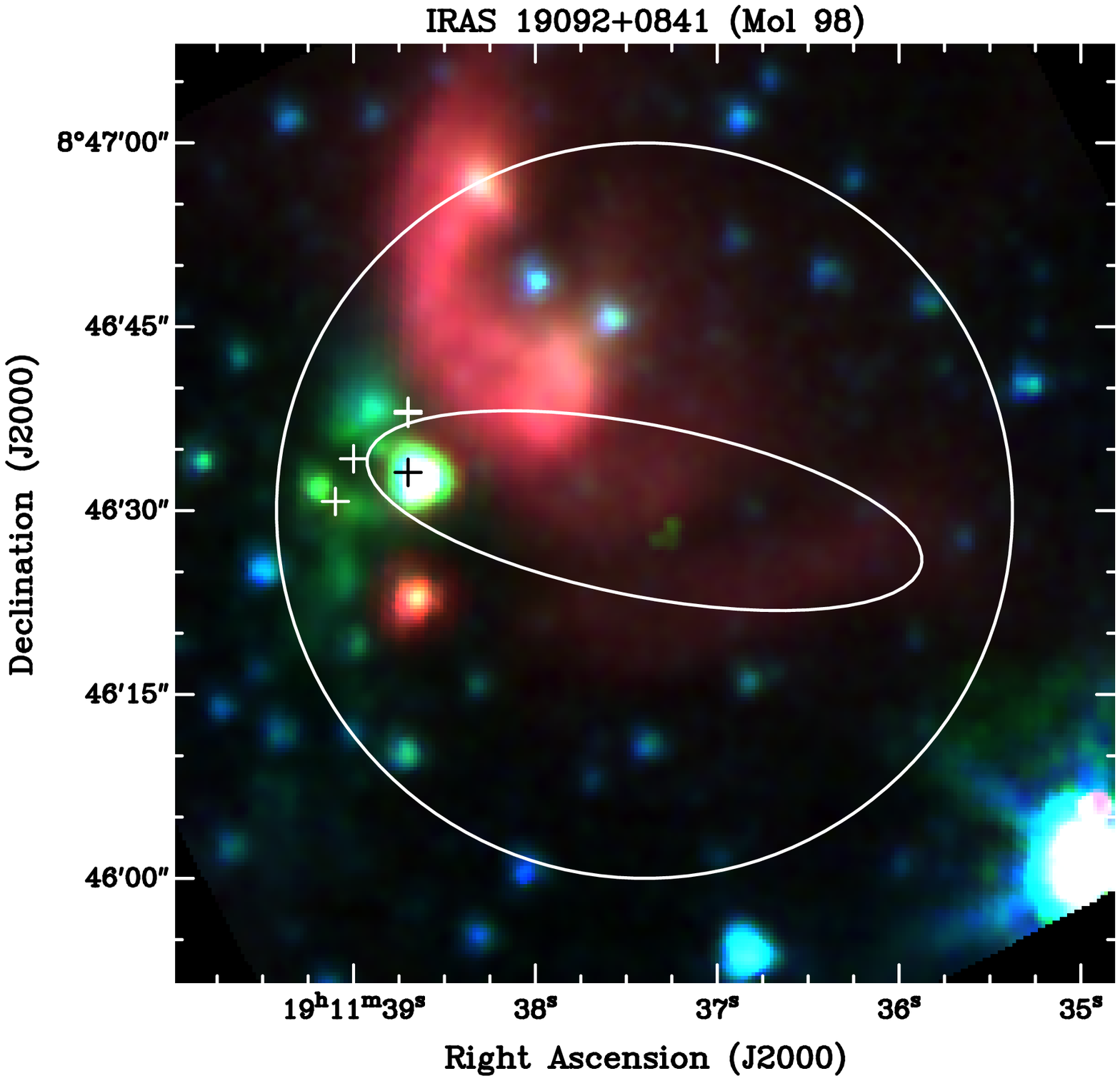} &
\includegraphics[width=.47\textwidth,angle=-90]{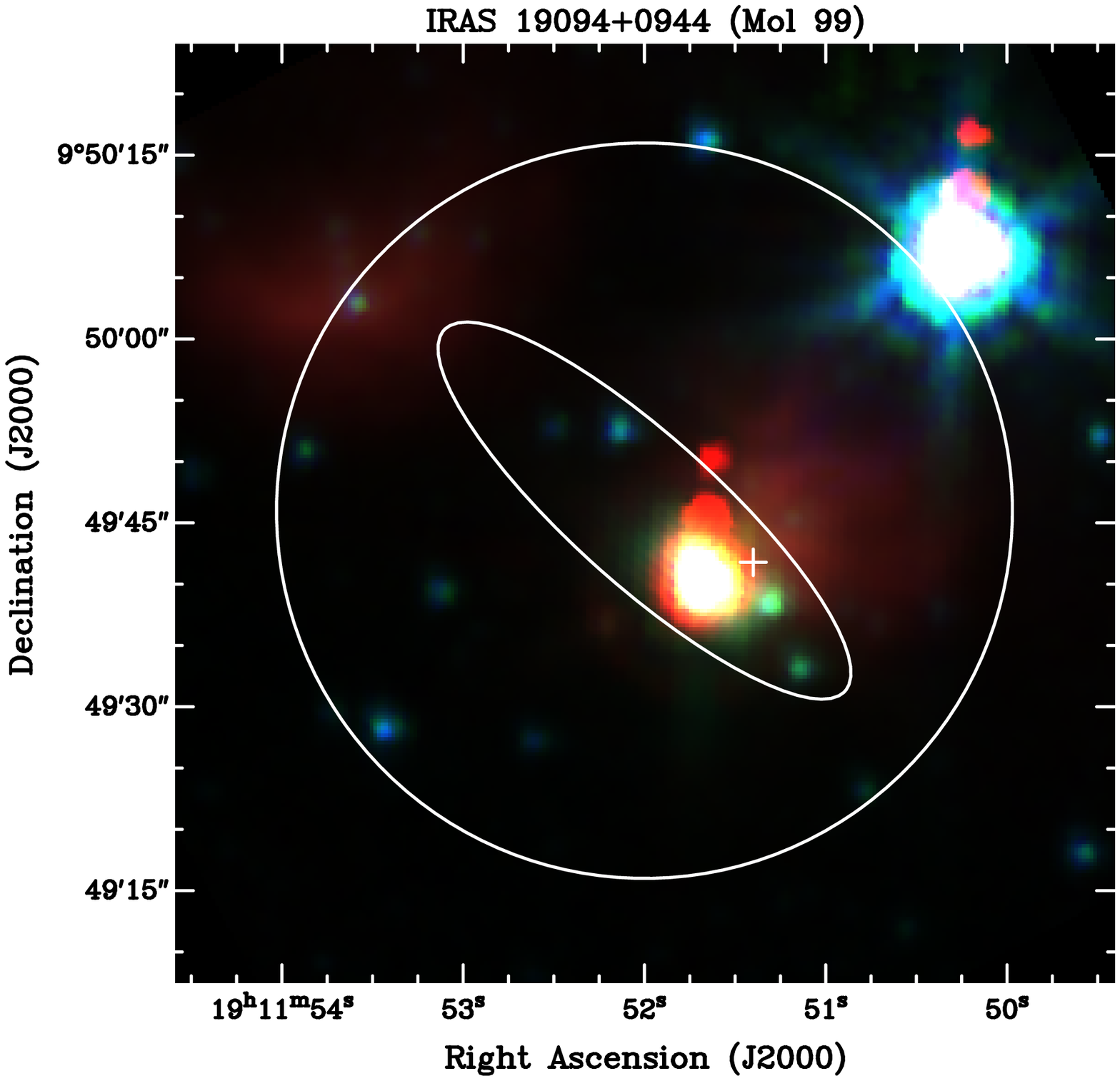} \\
\end{tabular}
\caption{Cont.
\label{3col-4}}
\end{figure}

\addtocounter{figure}{-1}
\begin{figure}
\epsscale{0.85}
  \begin{tabular}{@{}cc@{}}
\includegraphics[width=.47\textwidth,angle=-90]{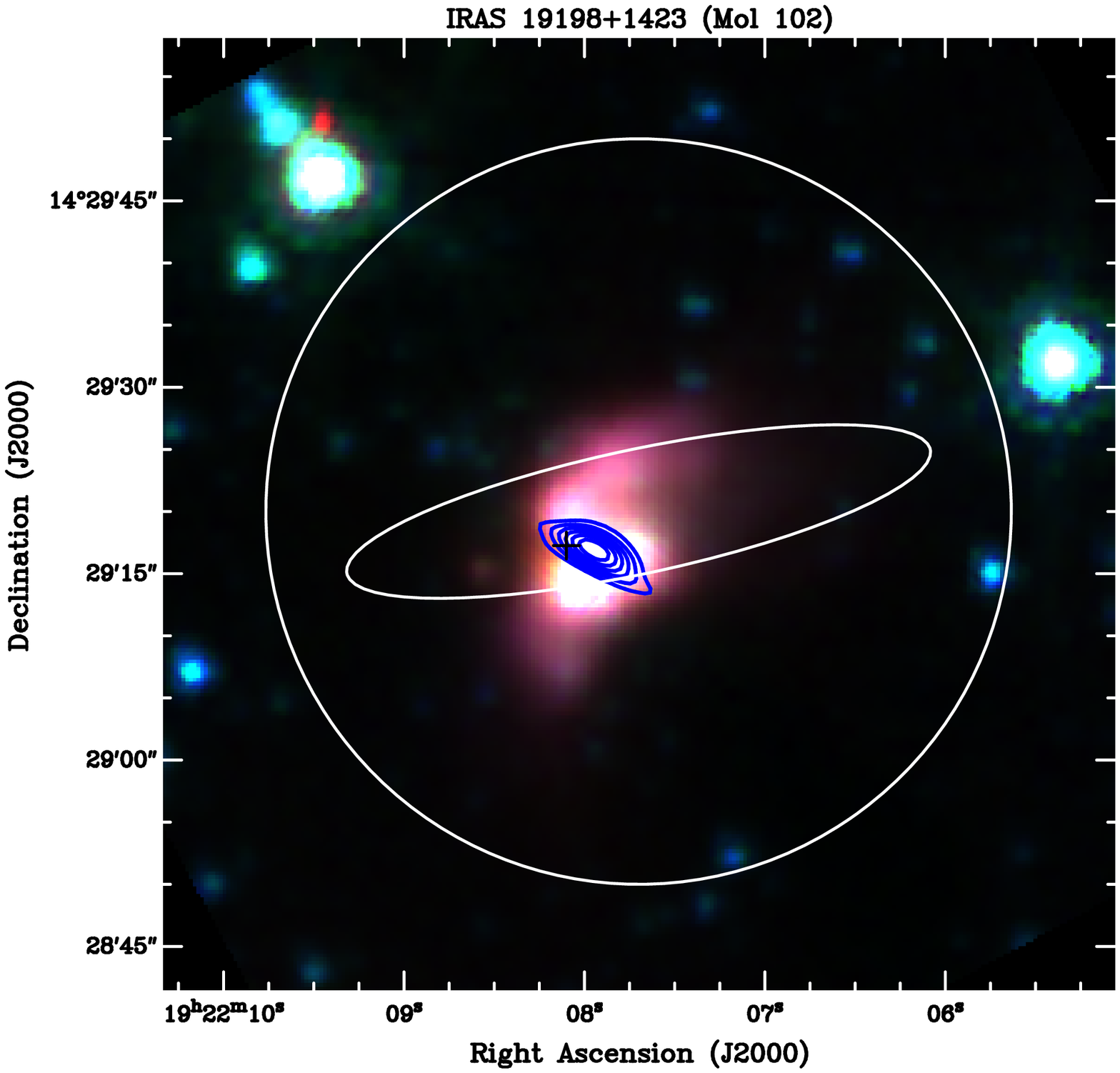} & 
\includegraphics[width=.47\textwidth,angle=-90]{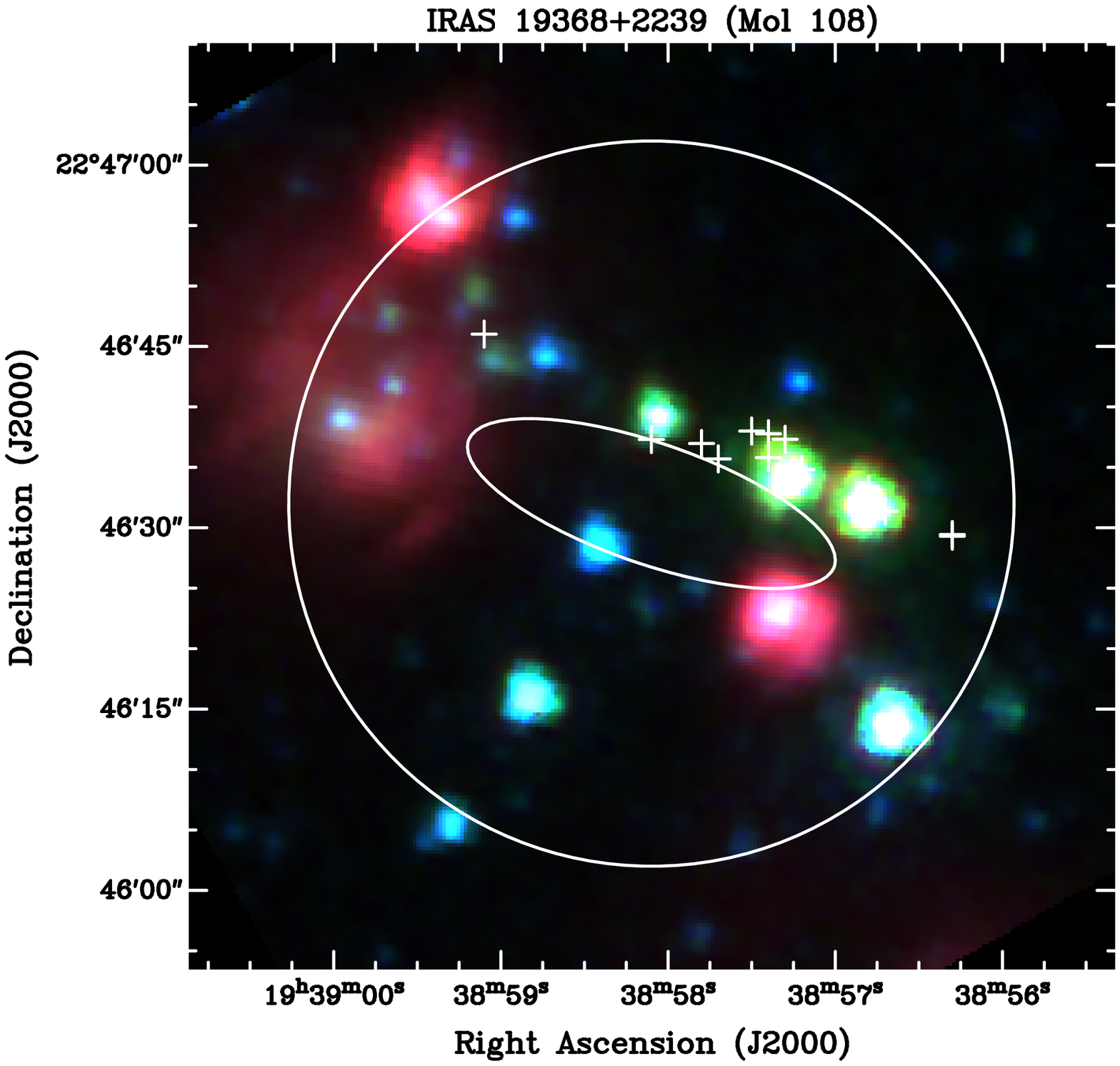} \\
\includegraphics[width=.47\textwidth,angle=-90]{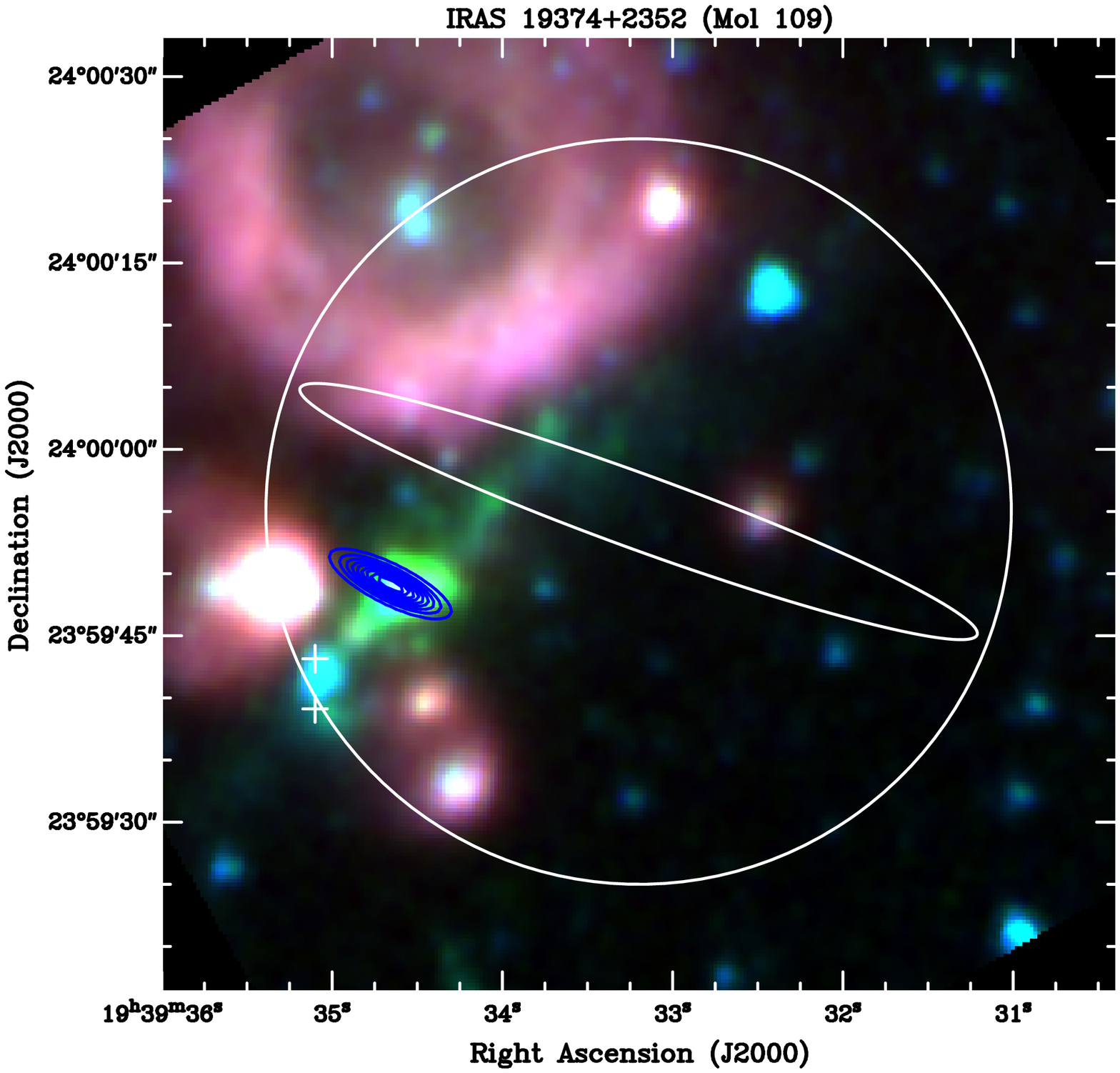} &
\includegraphics[width=.47\textwidth,angle=-90]{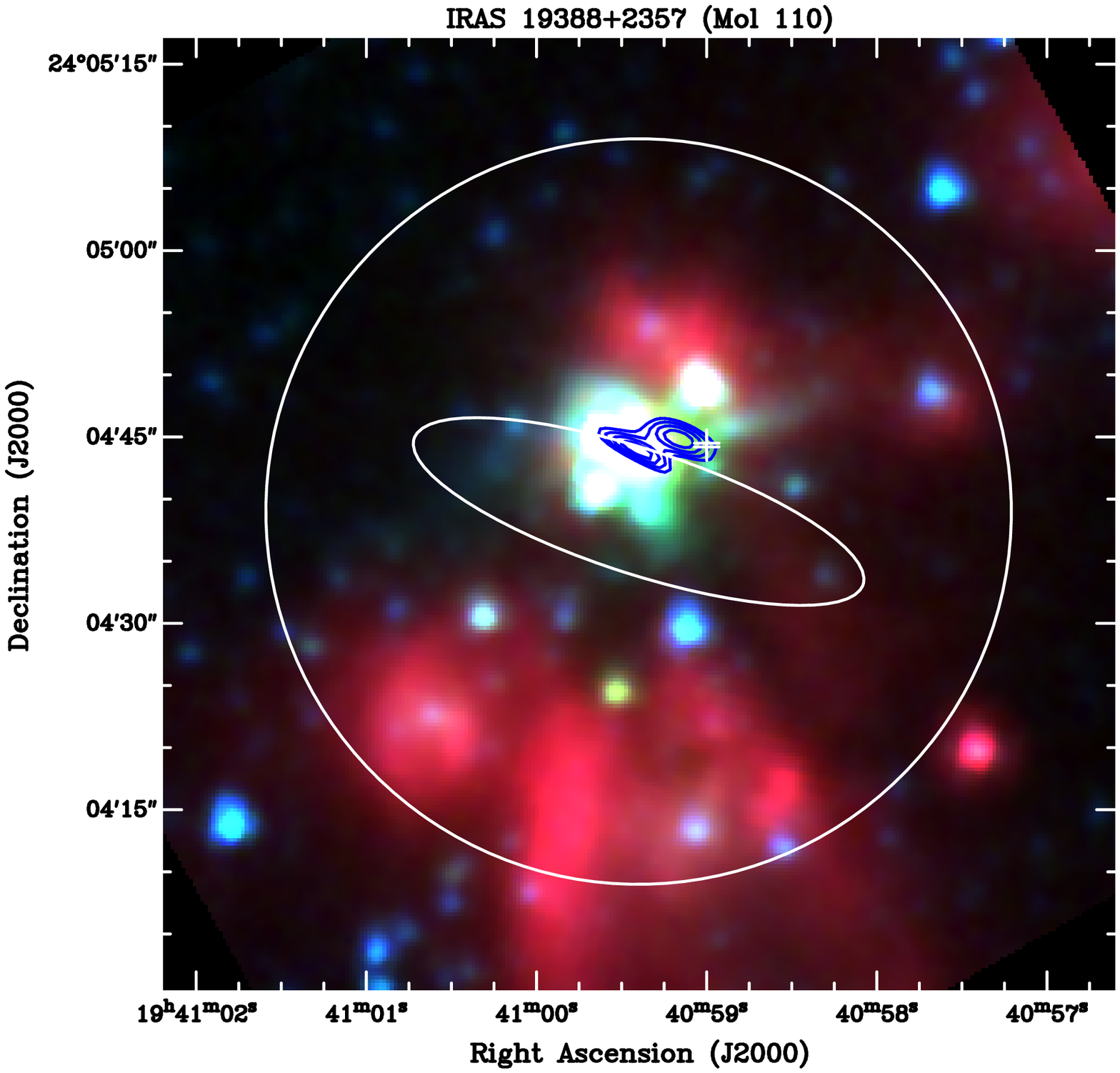} \\
\end{tabular}
\caption{Cont.
\label{3col-5}}
\end{figure}

\addtocounter{figure}{-1}
\begin{figure}
\epsscale{0.85}
  \begin{tabular}{@{}cc@{}}
\includegraphics[width=.47\textwidth,angle=-90]{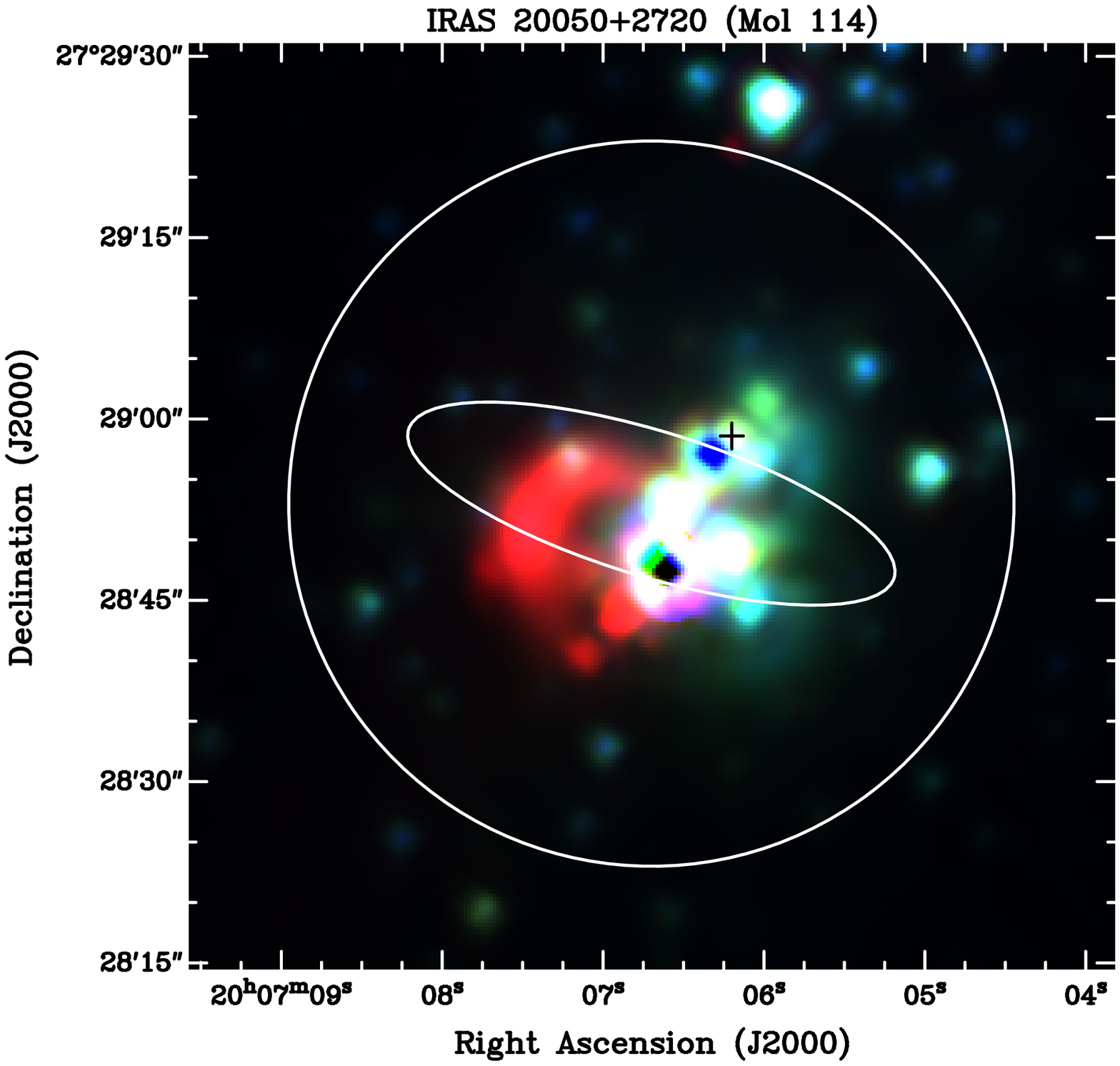} & 
\includegraphics[width=.47\textwidth,angle=-90]{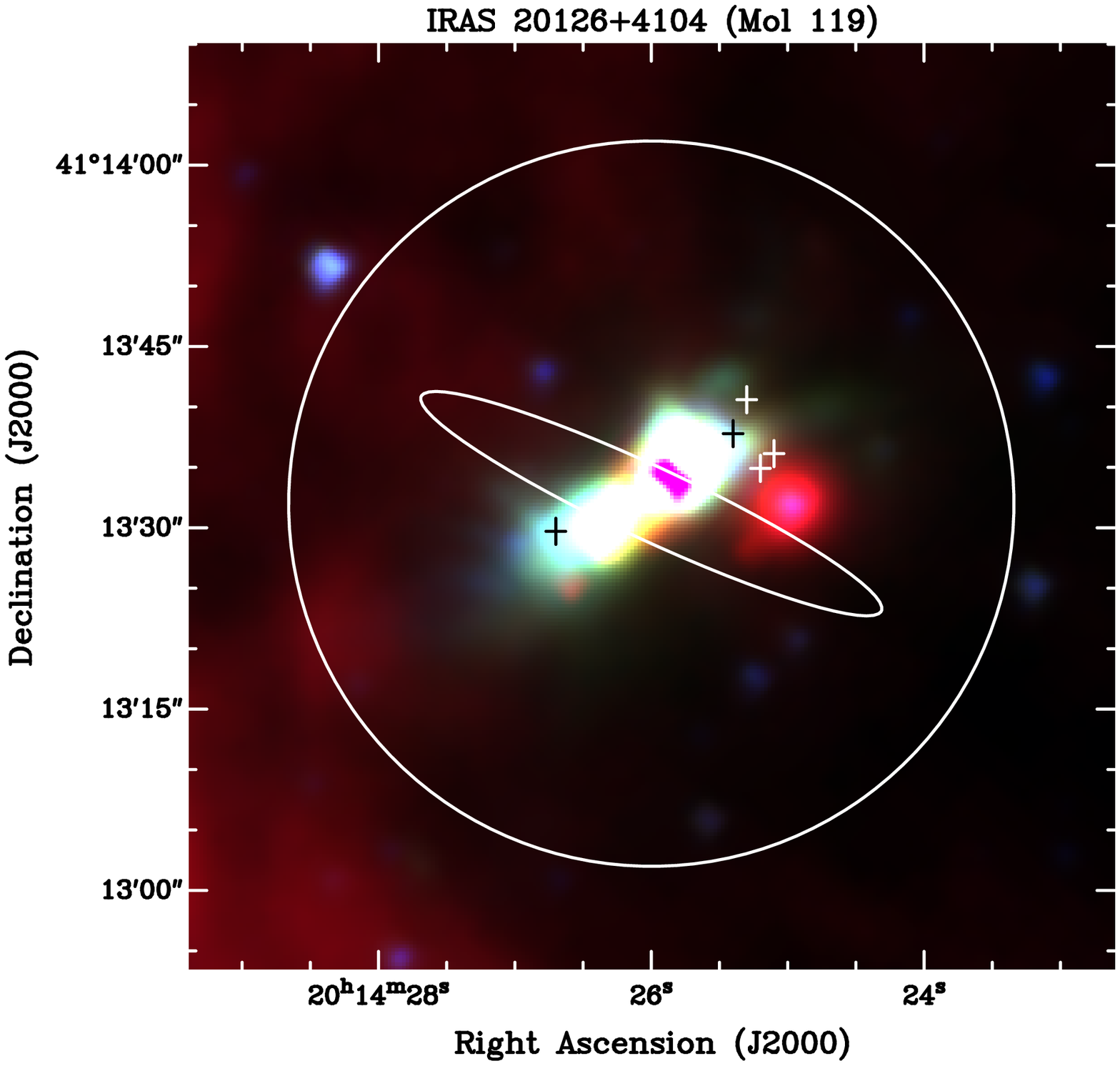} \\
\includegraphics[width=.47\textwidth,angle=-90]{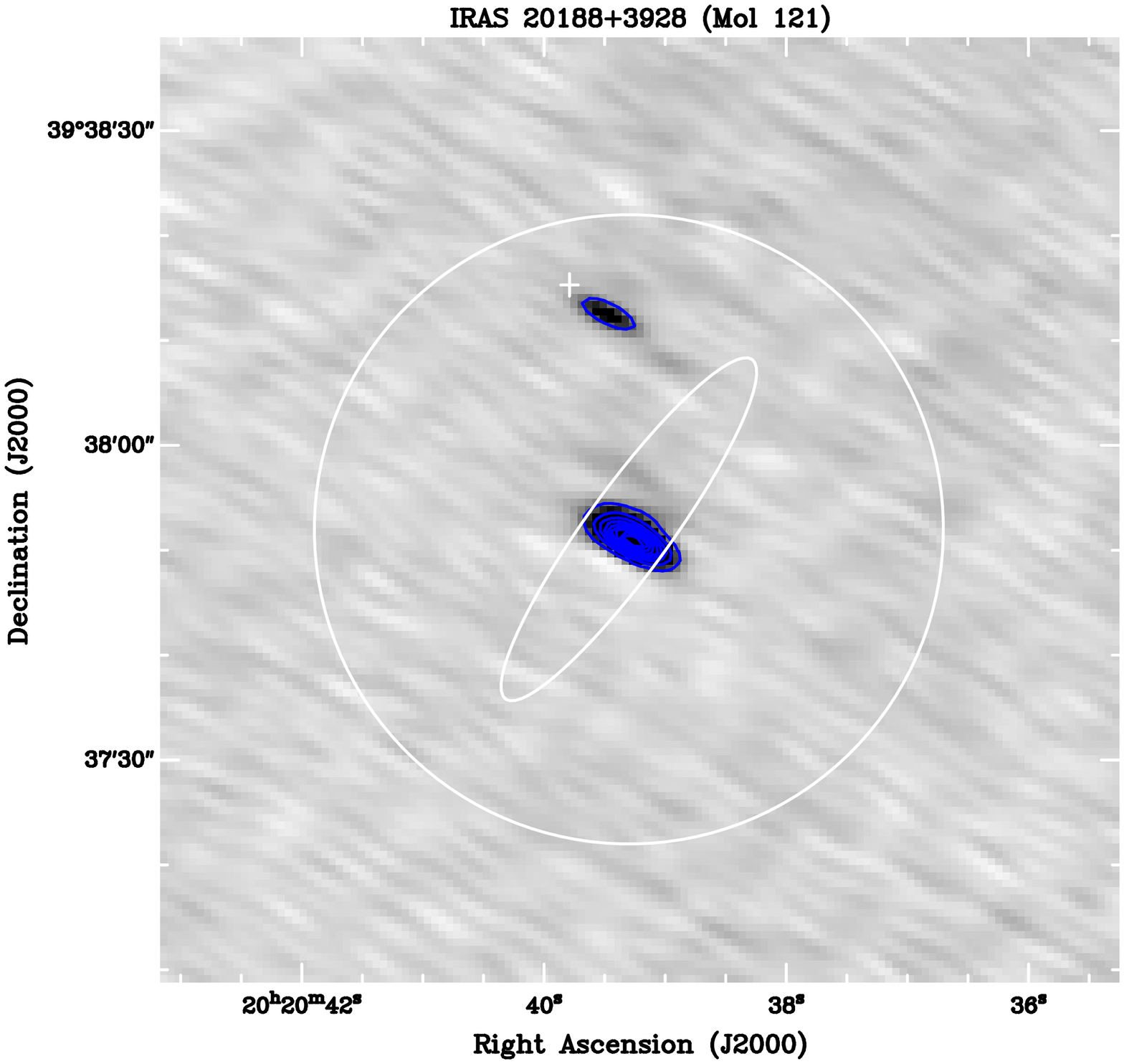} &
\end{tabular}
\caption{Cont.
\label{3col-6}}
\end{figure}

\end{document}